\documentclass[AMA,Times1COL]{wileyNJDv5} %STIX1COL,STIX2COL,STIXSMALL

\articletype{Article Type}%

\received{Date Month Year}
\revised{Date Month Year}
\accepted{18 January 2025}
\journal{Statistics in Medicine}
\volume{00}
\copyyear{2025}
\startpage{1}

\usepackage{mathtools}

\begin{document}

\title{A Calibrated Sensitivity Analysis for Weighted Causal Decompositions}

\author[1]{Andy A. Shen}
\author[2]{Elina Visoki}
\author[2,3]{Ran Barzilay}
\author[1]{Samuel D. Pimentel}

\authormark{SHEN \textsc{et al.}}
\titlemark{Sensitivity Analysis for Decompositions}

\address[1]{\orgdiv{Department of Statistics}, \orgname{University of California, Berkeley}, \orgaddress{\state{California}, \country{USA}}}

\address[2]{\orgdiv{Children's Hospital of Philadelphia}, \orgaddress{\state{Pennsylvania}, \country{USA}}}

\address[3]{\orgdiv{Perelman School of Medicine}, \orgname{University of Pennsylvania}, \orgaddress{\state{Pennsylvania}, \country{USA}}}

\corres{Andy Shen, PhD Student, 367 Evans Hall, University of California
Berkeley, CA 94720-3860. \email{aashen@berkeley.edu}}

%\fundingInfo{Text}
%\JELinfo{ejlje}

\abstract[Abstract]{Disparities in health or well-being experienced by minority groups can be difficult to study using the traditional exposure-outcome paradigm in causal inference, since potential outcomes in variables such as race or sexual minority status are challenging to interpret. Causal decomposition analysis addresses this gap by positing causal effects on disparities under interventions to other, intervenable exposures that may play a mediating role in the disparity. While invoking weaker assumptions than causal mediation approaches, decomposition analyses are often conducted in observational settings and require uncheckable assumptions that eliminate unmeasured confounders. Leveraging the marginal sensitivity model, we develop a sensitivity analysis for weighted causal decomposition estimators and use the percentile bootstrap to construct valid confidence intervals for causal effects on disparities. We also propose a two-parameter reformulation that enhances interpretability and facilitates an intuitive understanding of the plausibility of unmeasured confounders and their effects. We illustrate our framework on a study examining the effect of parental support on disparities in suicidal ideation among sexual minority youth. We find that the effect is small and sensitive to unmeasured confounding, suggesting that further screening studies are needed to identify mitigating interventions in this vulnerable population.}


% \jnlcitation{\cname{%
% \author{Taylor M.},
% \author{Lauritzen P},
% \author{Erath C}, and
% \author{Mittal R}}.
% \ctitle{On simplifying ‘incremental remap’-based transport schemes.} \cjournal{\it J Comput Phys.} \cvol{2021;00(00):1--18}.}

\maketitle

\renewcommand\thefootnote{}
%\footnotetext{\textbf{Abbreviations:} ANA, anti-nuclear antibodies; APC, antigen-presenting cells; IRF, interferon regulatory factor.}

\newenvironment{revisiontwo}{\color[HTML]{000000}}{\ignorespacesafterend}
\newcommand{\smallrevisiontwo}[1]{\textcolor[HTML]{000000}{#1}}

\newenvironment{revisionone}{\color[HTML]{000000}}{\ignorespacesafterend}
\newcommand{\smallrevisionone}[1]{\textcolor[HTML]{000000}{#1}}

%% Andy macros
\newcommand\responsetwo[1]{\;
{\textcolor[HTML]{000000}{{\bf{Response}: #1} \;}
}} 

\newcommand\responseone[1]{\;
{\textcolor[HTML]{000000}{{\bf{Response}: #1} \;}
}} 

\renewcommand\thefootnote{\fnsymbol{footnote}}
\setcounter{footnote}{1}
\section{Introduction} \label{sec:introduction}
Sexual minority youth (identifying as lesbian, gay, or bisexual) face significantly higher risks of adverse mental health burden than their heterosexual (non sexual minority) peers, placing them at a severe disadvantage in society.\citep{feinstein2024sexual,gordon2024role,nagata2023sexual} A major consequence of poor mental health is an increase in suicide risk, which is among the top three leading causes of death for adolescents and young adults in the United States.\citep{wisqars2024cdc} A meta-analysis of 20 studies found that 28\% of sexual minority youth reported a history of suicidality (suicidal ideation or behavior) compared to 12\% of heterosexual youth.\citep{marshal2011suicidality} Sexual minority youth are at least 2.9 times more likely to experience suicidal ideation compared to heterosexual youth, even after adjusting for risk factors such as discrimination and structural stigma of sexual minorities.\citep{gordon2024role} To address these risk disparities, there is a critical need to develop interventions that mitigate suicide risk among sexual minority individuals.\citep{melvin2024scoping} Recent studies have indicated promising results \citep{russon2021implementing} and research from randomized controlled trials (RCTs) in other high risk populations have also reported that suicidal ideation can be reduced through interventions such as online self-help programs and pharmacotherapy.\citep{van2018effectiveness,ballard2021clinical} The remaining challenge is to identify interventions that reduce the most disparity in suicidal ideation for sexual minority youth.

Leveraging observational data helps prioritize intervention targets given the high costs associated with generating evidence from an RCT. In practice, the effectiveness of hypothesized intervention targets is \smallrevisionone{evaluated} using \textit{causal decomposition analysis} (or ``causal decompositions'').\citep{jackson2018decomposition,jackson2021meaningful} Suppose researchers hypothesize that disparities in \textit{parental support} between sexual minority and heterosexual youth may have an effect on suicidality. While parental support has been linked to lower suicidality in sexual minority youth \citep{delferro2024role}, causal decompositions provide needed transparency to these analyses by constructing treatment effects with respect to an intervention. Here we are interested in the suicidal ideation rate for sexual minorities in a counterfactual world where their levels of parental support are equivalent to that of heterosexual youth. This counterfactual measurement breaks down a descriptive disparity into two terms: a \textit{disparity reduction} term characterizing the portion of the disparity that is reduced through the target intervention and a \textit{residual disparity} term characterizing how much disparity remains after equalizing parental support across the two groups \citep{vanderweele2014causal,jackson2018decomposition,jackson2021meaningful}:
\begin{align*}
    \mathrm{Observed~Disparity}~=~ \mathrm{Disparity~Reduction}~+~ \mathrm{Residual~Disparity}.
\end{align*}
Researchers use causal decompositions to identify factors that account for a large part of the disparity, which determines their potential as intervention targets. In order to learn about interventional effects from observational data, observed confounders must be controlled for and unobserved confounders must be absent or limited in impact. Accordingly, a \textit{sensitivity analysis} is performed to determine the degree of unmeasured confounding necessary to significantly alter or reverse a study's results. These analyses summarize uncertainty to potential omitted confounders, providing researchers with a transparent platform to reason about whether confounders of such strength could exist in their study. Motivated by the hypothesis of parental support as a target intervention to mitigate suicidal ideation risk in sexual minority youth, we develop a sensitivity analysis framework for weighted causal decomposition estimators.

\subsubsection{Methodological contributions} We make several important contributions in causal decomposition analysis and sensitivity analysis. Existing sensitivity analysis methods for causal decompositions are primarily tailored to regression-based frameworks while our framework focuses specifically on \textit{weighted estimators} for causal decompositions. Weighting estimators are advantageous since they protect against extrapolation and post-selection inference issues from specification searching. We adopt the \textit{marginal sensitivity model (MSM)} and we prove that, under mild assumptions, the percentile bootstrap yields valid $1-\alpha$ confidence intervals for the causal estimand of interest. To the best of our knowledge, our sensitivity framework is the first to provide asymptotically valid confidence intervals for inference under unobserved confounding in causal decomposition analysis. We also offer a reformulation of the MSM, which re-parameterizes a one-dimensional sensitivity analysis into two dimensions. This serves as a practical tool for practitioners to easily calibrate and interpret their sensitivity analysis results with minimal assumptions.

\subsubsection{Substantive contributions} Our methodological contributions provide key capabilities for assessing target interventions that mitigate group disparities. To demonstrate our sensitivity framework, we leverage data from the Adolescent Brain Cognitive Development (ABCD) Study \citep{karcher2021abcd}, a longitudinal study of youth behavior in the United States (see Section \ref{sec:running-example} and Section C of the Appendix). Specifically, we assess the impact of parental support on disparities in suicidal ideation among sexual minority youth and account for unobserved confounding. Our analysis indicates that intervening on parental support is beneficial but yields only modest disparity reductions in suicidal ideation. The observed disparity reduction is also sensitive to unmeasured confounding, suggesting that other interventions may effectively address suicidal ideation in this vulnerable population and highlighting the importance of sensitivity analysis tools like ours. Our findings add important nuance to existing work on parental support and suicidality in sexual minority youth which has generally ignored unobserved confounding and has relied on parametric linear modeling. \citep{delferro2024role} \newline 

\noindent This article is organized as follows. In Section \ref{sec:running-example}, we discuss the ABCD dataset in more detail. Section \ref{sec:background} describes the notational framework of causal decompositions and the corresponding sensitivity analysis along with a review of these areas. In Section \ref{sec:sens-model-paper}, we discuss our sensitivity framework and show that the percentile bootstrap yields valid confidence intervals for the causal estimand of interest. Section \ref{sec:amplification} introduces our reformulation and discusses its benefits. We demonstrate our sensitivity analysis and its reformulation on the ABCD dataset in Section \ref{sec:application} and conclude in Section \ref{sec:discussion}.

\subsection{The Adolescent Brain Cognitive Development (ABCD) Study} \label{sec:running-example}

We utilize data from the Adolescent Brain Cognitive Development (ABCD) Study. This study enrolled participants ($N =11,868$) ages nine to ten at 21 research sites across the United States through school-based recruitment. \citep{garavan2018recruiting} The aim of the ABCD Study is to elucidate mechanisms that contribute to risk and resilience in development of brain and behavior. Study participants are assessed annually and the study protocol involves deep characterization of participants’ mental health, neurocognitive profiles, and environment and lifestyle factors. Assent was obtained from all participants and parents/guardians gave written informed consent. The ABCD Study protocol was approved by the University of California San Diego Institutional Review Board. The current study was exempted from a full review by the University of Pennsylvania Institutional Review Board. We specifically use longitudinal data from ABCD Study Data Release 5.1 including measures collected at baseline, one-, two- and three-year assessment waves. \smallrevisiontwo{The ABCD Study began in 2016 and our analysis uses data collected during the period 2016-2022.} We generated binary variables for the intervention/exposure (parental support) and outcome (suicidal ideation) for each participant in line with previous work. \citep{barzilay2022genetic} The analyses included 11,622 ABCD Study participants and do not include 254 (2.1\%) participants who had missing data for sexual minority identity at all time points. As mentioned in Section 1.1.1, participants who did not understand or declined to answer the question about sexual identity were treated as missing. This was done across the four ABCD Study assessments.

\subsubsection{Definition of sexual minority} As part of the study protocol, participants were asked about their sexual identity, allowing researchers to study contributions to risk and resilience in sexual minority youth and to examine disparities between sexual minority and heterosexual youth. In accordance with Gordon et al. (2024) \cite{gordon2024role}, sexual minority identity ($G$) was determined using the question ``Are you gay or bisexual?'' with possible responses of ``yes,'' ``no,'' ``maybe,'' ``I do not understand,'' or ``Refuse to answer''. We constructed a binary variable for sexual minority status (``yes'' and ``no''). Participants who answered ``yes'' or ``maybe'' to the above question were classified as having sexual minority identity ($G=1$) and participants who answered ``no'' to the above question were classified as having heterosexual (non sexual minority) identity ($G=0$). Participants who responded ``I do not understand'' or ``decline to answer'' were treated as missing. This was done across the four ABCD Study assessments. \smallrevisiontwo{To remain consistent with previous analyses on sexual minorities using the ABCD study \cite{gordon2024role}, our definition of sexual minority does not include youth who identified as transgender (asked in a separate question) or asexual (not asked). Effects on these subpopulations are an important topic for future work.} 

\subsubsection{Definition of parental support} Our target intervention of ``parental support'' was determined based on the youth’s perception of how their parents could comfort and support them through difficult times. This exposure was chosen based on literature suggesting that parental support and support is associated with less suicidality among sexual minority youth, including those in the ABCD study. \citep{katz2016lesbian,klein2023mediating} As part of the study protocol, participants were asked a series of questions about their parents and/or caregivers. Our definition of parental support uses responses to the following questions:
\begin{itemize}
    \item {My parent/caregiver makes me feel better after talking over my worries with him/her.}
    \item {My parent/caregiver is able to make me feel better when I am upset.}
\end{itemize}
Participants were asked to rate how much they agreed with each question on a scale from 1 to 3, where each number corresponds to the following description about the parent/caregiver: 
\begin{itemize}
    \item 1 - Not like him/her
    \item 2 - Somewhat like him/her
    \item 3 - A lot like him/her
\end{itemize}
parental support ($Z$) was binarized using the following threshold: we assigned $Z=1$ (superior parental support, exposed group) to for youth who rated 3 to \textit{both} questions for \textit{all} parents/caregivers. Conversely, we assigned $Z=0$ (poor parental support, unexposed group) to youth who did not respond to \textit{any} of the two questions for \textit{any} parent/caregiver with a rating of 3. In other words, $Z=1$ only if all ratings for all parents/caregivers were 3 and $Z=0$ if no rating for any parent/caregiver was 3. After this binarization, we were left with $N = 4510$ children (out of 11622) for the final analysis. \smallrevisiontwo{From this cohort, the age range at the start of the third assessment wave for sexual minorities was 11.58 - 14.42 years, and the age range for heterosexual youth was 11.42 - 14.58 years.}

\subsubsection{Measurement of suicidal ideation} The clinical assessment of participants in the ABCD Study included a deep characterization of suicidal thoughts and behavior based on the self-report using the validated and computerized Kiddie-Structured Assessment for Affective Disorders and Schizophrenia for DSM-5 (KSADS-5). \citep{kaufman1997schedule} The KSADS-5 assessed the following symptoms: passive suicidal ideation, active but non-specific suicidal ideation, suicidal ideation with a specific method, active suicidal ideation with an intent, active suicidal ideation with a plan, preparatory actions toward imminent suicidal behavior, interrupted suicidal attempts, aborted suicidal attempts, and suicide attempts. \citep{janiri2020risk} As the proportion of suicide attempts were low, we focus the current analysis on suicidal ideation in any of the forms described above, and we collapsed all suicidal ideation items into a binary measurement that we refer to as ``suicidal ideation''.

\subsubsection{Choice of covariates} Since youth suicide is a complex behavior that is influenced by many psychological factors in addition to parental support \citep{carballo2020psychosocial}, we compiled a list of potential confounders that we include as covariates. These covariates include age, sex assigned at birth, sibling order and number of siblings, income, family conflict, peer victimization, school safety, neighborhood safety, neighborhood area deprivation index (ADI), and structural stigma against sexual minorities (state-level). The covariates were chosen to represent different facets of an individual's psychosocial environment in line with the \textit{ecological system theory}. This idea highlights how different layers of environment contribute to human development, including familial, school, neighborhood and wider societal environment. \citep{bronfenbrenner2005making} We discuss the selection and inclusion of these covariates in more detail in Section \ref{sec:abcd-initial}.

\section{Background} \label{sec:background}
\subsection{Setup and Notation} \label{sec:setup}
Consider an observational study of $n$ individuals indexed from $i = 1, \dots ,n$. As defined in Section \ref{sec:running-example}, let $G_i \in \bracesbinary$ represent sexual minority status and $Z_i \in \bracesbinary$ represent parental support. Since $G_i$ represents an inherent characteristic of an individual, it is not modifiable and cannot ``cause'' an effect. \citep{rubin1974estimating,holland1986statistics} More generally, $G_i$ is referred to as the \textit{`group'} or \textit{`characteristic'}. On the other hand, $Z_i$ is a manipulable exposure since there exists interventions (such as an encouragement design) that alter one's level of parental support. We refer to $Z_i$ as the \textit{`exposure'} or \textit{`intervention'}, where $Z_i = 1$ if unit $i$ receives the intervention and $Z_i = 0$ otherwise. Each individual also has a vector of background covariates\footnote{The background covariates may be partitioned into different subgroups based on ethical and moral grounds; see Section \ref{sec:allowability-background} for further information.} $X_i \in \mathcal{X} \subset \R^{d}$ and outcome $Y_i \in \R$. We denote the unobserved confounder as $U_i \in \mathbb{R}$. Since the exposure $Z$ is modifiable, we can also define potential outcomes using the traditional exposure-outcome paradigm:
\begin{align*}
    Y_i = Z_i Y_i(1) + (1-Z_i) Y_i(0),
\end{align*}
where $Y_i(z)$ denotes the potential outcome of unit $i$ when $Z_i=z$. \smallrevisionone{Our formulation relies on the Stable Unit Treatment Value Assumption (SUTVA) \citep{rubin1974estimating}, which posits one version of the intervention across all units and no interference in the intervention assignment process. We assume the tuples $\paren{G_i, Z_i, X_i, U_i, Y_i(1), Y_i(0)}$ are sampled} independently from  \smallrevisionone{a single} distribution $\P$ and drop the subscript $i$ where convenient. We also assume the covariates $X_i$ and sexual minority status $G_i$ all temporally occur before parental support $Z_i$ and suicidal ideation $Y_i$; see Section B of the Appendix for more discussion.

We define $\mu_g \coloneqq \E\brackets{Y \mid G=g}$ as the expected outcome for those in group $g$, estimated by the sample mean\footnote{For other choices of estimands to define group-wise outcomes, see Jackson (2021).\cite{jackson2021meaningful}}. This allows us to quantify the \textit{observed disparity} $\tau$ as 
\begin{align}
    \label{eqn:obs-disparity}
    \tau \coloneqq \mu_1 - \mu_0.
\end{align}
$\tau$ is given a causal interpretation in traditional causal inference settings because it is measured with respect to an intervenable exposure. However, $\tau$ is non-causal in the causal decomposition setting since it is measured with respect to the immutable variable $G$. 

The focus of causal decomposition analysis is to measure how an observed group disparity would change if we equalized the distribution of intervention across groups. Let $R_{Z \mid G=0, x} \in \left\lbrace 0,1 \right\rbrace$ denote a random draw from $P\paren{Z \mid G = 0, X = x}$. \citep{jackson2021meaningful,park2023sensitivity,yu2023nonparametric} This functions as a stochastic intervention that aligns the conditional distribution of $Z$ for group $G = 1$ with that of group $G = 0$ for those with the same covariates $X=x$. Our target estimand is the mean counterfactual outcome for group $G=1$ after each member of this group with covariates $x$ has received treatment according to the distribution of $R_{Z \mid G=0, x}$:
\begin{align}
    \muint \coloneqq \E\brackets{Y\left(R_{Z \mid G=0, x}\right) \mid G = 1}.
\end{align}
In our application, $\muint$ measures the average suicidal ideation rate for sexual minorities if their distribution of parental support was set to \smallrevisionone{that under} the same level of heterosexuals with the same baseline covariates $x$. For ease of interpretation, we use the notation $Y(int)$ for $Y\left(R_{Z \mid G=0, x}\right)$ to reflect that the outcome is being measured with respect to a counterfactual stochastic intervention.

Identification of $\muint$ facilitates decomposing $\tau$ into two distinct \textit{causal estimands}:
\begin{align}
    \label{eqn:disparity-decomp}
    \tau = \underbrace{\mu_1 - \muint}_{\mathrm{disparity~reduction}} + \underbrace{\muint - \mu_0.}_{\mathrm{residual~disparity}}
\end{align}
The first component, $\mu_1 - \muint$, represents the \textit{disparity reduction} for group $G=1$, reflecting the expected \smallrevisionone{change} in suicidal ideation rate in the sexual minority group when their distribution of parental support is \smallrevisionone{changed to be} equivalent to that of the heterosexual group. \smallrevisionone{Intuitively, one may think of this as the proportion of sexual minority youth who experienced suicidal ideation but would not have if their parental support had been improved.} The second term, $\muint - \mu_0$, is the \textit{residual disparity}, which is the difference in suicidal ideation between the two groups that remains after equalizing \smallrevisionone{the distribution of} parental support across the two groups via $R_{0}$. More concretely, this estimand measures the disparity that is not explained after equalizing the distribution of parental support. Similar values of $\mu_1$ and $\muint$ suggest that $Z$ does not contribute \smallrevisionone{to} the observed disparity. \smallrevisionone{When $\mu_1$ and $\muint$ are different but $\mu_0$ and $\muint$ are similar,} then the treatment $Z$ \smallrevisionone{may} reflect an effective intervention worthy of consideration by policymakers and other stakeholders. %\smallrevisionone%{However, interpreting the potential efficacy of $Z$ may be more challenging if the observed disparity $\tau$ is small to begin with.}

Identification of causal parameters in observational studies assumes that conditioning on observed covariates $X$ sufficiently removes any confounding bias between exposure and outcome. We consider a similar version of this assumption in causal decomposition settings:
\begin{assumption}[Conditional ignorability among $G=1$] \label{as:condig-paper}
    \begin{align*}
        Y(z) \indep Z \mid G = 1, X
    \end{align*}
    for all $z \in \bracesbinary$.
\end{assumption}
This assumption states that the potential outcome for sexual minorities is independent from their parental support conditional on all observed covariates among sexual minorities\footnote{This assumption is not necessary for individuals in group $G=0$ since we only consider the stochastic intervention on those in group $G=1$.}. We consider violations to this assumption in our sensitivity analysis.

Next, define the \textit{group propensity scores} as follows:
\begin{align}
    \label{eqn:group-propensity}
    e_1(X) &\coloneqq  P\paren{Z = 1 \mid G = 1, X = x}\\
    e_0\paren{X} &\coloneqq P\paren{Z = 1 \mid G = 0, X=x}.
\end{align}
For simplicity, we will often use $e_0$ and $e_1$ in place of $e_0(X)$ and $e_1({X})$, respectively. This motivates the \textit{overlap in treatment assumption}:
\begin{assumption}[Overlap in treatment assignment] \label{as:overlap-paper}
    \begin{align*}
        0 < e_g < 1 \quad\text{for all } g \in \bracesbinary.
    \end{align*}
\end{assumption}
Under Assumption \ref{as:overlap-paper}, the probability of having superior parental support for both sexual minority and heterosexual youth is nonzero. Assumption \ref{as:overlap-paper} is analogous to the overlap condition of propensity scores in standard inverse propensity score weighting analyses, which is necessary for estimand identification. In the causal decomposition setting, both group propensity scores must satisfy the overlap condition to identify $\muint$.

Under Assumptions \ref{as:condig-paper} and \ref{as:overlap-paper}, we can identify $\muint$ as
\begin{align} \label{eqn:rmpw-estimand}
    \muint = \E\brackets{w Y\mid G = 1} = \E\brackets{\paren{\wrmpw} Y\mid G = 1},
\end{align}
where $w = w(X, Z) \coloneqq  \wrmpw$ is the \textit{ratio of mediator probability weight} (RMPW). \citep{hong2010ratio,hong2018weighting,hong2021did,jackson2021meaningful} See Section A.1 in the Appendix and Jackson (2021) \cite{jackson2021meaningful} for a full derivation. 

Equation \eqref{eqn:rmpw-estimand} can be estimated using a weighted sample average of the outcomes in group $G = 1$:
\begin{align}
    \label{eqn:rmpw-estimator}
    \hat{\mu}_{R_0} = \frac{\sumin G_i \hat{w}_i Y_i}{\sumin G_i \hat{w}_i},
\end{align}
where $\hat{w}_i$ is the estimated RMPW weight for unit $i$. 
In general, the two propensity scores used to estimate $\hat{w}_i$ are model-agnostic. However, in Section \ref{sec:confint-paper} we consider the special case when $e_0$ and $e_1$ follow a \smallrevisiontwo{parametric} logistic model and derive inferential guarantees for the percentile bootstrap procedure, demonstrating its validity as a valid $1-\alpha$ confidence interval.

\subsection{Disparity estimation under allowability frameworks} \label{sec:allowability-background}
The decomposition introduced in the previous section facilitates a broad understanding of the target intervention by considering how \textit{all} covariates contribute to mitigating a disparity. Researchers instead may be interested in more nuanced insights where a target intervention is tailored within levels of \textit{certain} covariates. For instance, the intervention of parental support may be applied differently for boys and girls or for pre-adolescent and teenage youth. However, indiscriminately tailoring the intervention within levels of all measured covariates may not be practical or morally acceptable. Consider the income covariate: Deploying a parental support intervention differently within levels of family income (i.e., adjusting for income) would mitigate disparities in parental support \textit{within} income brackets but not \textit{across} income brackets. This preserves income-related disparities in parental support which may inadvertently further disadvantage minority groups.

Therefore, it is important to consider the equity implications of adjusting for specific covariates in disparity measurement and estimation. The \textit{allowability framework} \cite{duan2008disparities,jackson2018decomposition,jackson2021meaningful,chang2024importance} delineates the considerations that arise in adjusting for certain covariates in disparity measurement. Under this framework, background covariates are categorized into allowable and non-allowable covariates: $X = (X^A, X^N)$. Allowable covariates are those whose statistical adjustment is deemed acceptable under medical or ethical considerations. In our application, we denote age and sex as allowable. Non-allowable covariates are those whose adjustment is considered unethical or inequitable. These covariates should not be controlled for, as doing so conceals inequalities between groups by removing a potentially problematic source of difference from consideration. Non-allowable covariates are denoted by $X^N$. In our application, non-allowable covariates are the remaining covariates introduced in Section \ref{sec:running-example} besides age and sex. The allowability-based decomposition mirrors the \textit{conditional decomposition} from Yu and Elwert (2023) \cite{yu2023nonparametric}, which isolates a set of pre-treatment covariates that tailors the intervention. For an in-depth discussion of the allowability framework, we refer the reader to the references listed at the end of this sentence. \cite{duan2008disparities,jackson2021meaningful,jackson2022observational,chang2024importance}

In light of these concerns, we consider a stochastic intervention that depends only on certain covariates, corresponding to the allowable covariates. \citep{jackson2021meaningful} We define the \textit{allowability stochastic intervention} $R_{Z \mid G = 0, x^A}$ as a random draw from the distribution $P(Z = 0 \mid G = 0, X^A = x^A)$. This differs from $R_{Z \mid G = 0, x}$ defined in Section \ref{sec:setup} which inherently treats all covariates as allowable. The resulting counterfactual estimand can be written as 
\begin{align} \label{eqn:rmpw-estimand-allow}
    \mu_{R_{0}^{a}}= \E\brackets{\paren{\wrmpwallow} Y\mid G = 1},
\end{align}
where $e_{0a} = P(Z = 1 \mid G=0, X^A)$. Notice that this estimand is identical to the estimand introduced in Equation \eqref{eqn:rmpw-estimand} with the exception being that the group $G=0$ propensity score $e_{0a}$ in Equation \eqref{eqn:rmpw-estimand-allow} conditions on allowable covariates only. Moreover, the propensity score $e_1$ is computed using allowable and non-allowable covariates in both equations; see Section A.2 in the Appendix for identification. We emphasize that our methodology does not require classifying confounders as allowable or non-allowable unless otherwise specified. In particular, $\muint$ and $\muintallow$ are interchangeable, as are $e_0$ and $e_{0a}$. The general form of our sensitivity model introduced in Section \ref{sec:sens-model-paper} applies to both allowable and non-allowable unmeasured confounders. Only Theorem \ref{thm:amp-bias-decomp} in Section \ref{sec:amp-deriv-paper} requires distinguishing between allowable and non-allowable unmeasured confounders.

\subsection{Unmeasured confounding and sensitivity analysis}
A sensitivity analysis allows for violations of Assumption \ref{as:condig-paper} by positing the existence of the unmeasured confounder $U$ and measuring how strong it must be to reverse a study's results. Such reversals can be characterized by bringing either the point estimate or confidence interval to a certain value (most commonly zero). Since causal decompositions measure the extent to which a target intervention mitigates an existing disparity, researchers are primarily concerned with the disparity reduction term and the strength of unmeasured confounding necessary to eliminate any observed disparity reduction. 

Suppose there exists an unmeasured confounder $U$ such that Assumption \ref{as:condig-paper} holds:
\begin{equation} \label{eqn:ignorability-with-U}
    Y(z) \indep Z \mid G = 1, X, U.
\end{equation}
We denote the \textit{ideal weight} $\wast = w^{\ast}\paren{X, Z, U}$ as a function of $X$, $Z$, and $U$. In particular, 
\begin{align}
    \label{eqn:ideal-w}
    w^{\ast} = \frac{e_0^{\ast}}{e_1^{\ast}} Z + \frac{1-e_0^{\ast}}{1-e_1^{\ast}} (1-Z),
\end{align}
where 
\begin{align*}
    e_1^{\ast}(X, U) &:=  \mathbb{P}\left(Z = 1 \mid G = 1, ~X = x, U = u\right)\\
    e_0^{\ast}(X, U) &:= \mathbb{P}\left(Z = 1 \mid G = 0, ~X = x, U = u\right)
\end{align*}
are the ideal propensity scores that adjust for the unmeasured confounder\footnote{Under the allowability framework, $e_{0a}^{\ast} = \mathbb{P}\left(Z = 1 \mid G = 0, ~X^A = x^A, U = u\right)$ if $U$ is allowable and $e_{0a}^{\ast} = e_{0a}$ otherwise; see Section \ref{sec:amp-deriv-paper} and Section A.4 in the Appendix for further discussion on allowable vs non-allowable unmeasured confounders.}. We refer to $\wast$ as an ideal weight since it guarantees the identifiability of $\muint$ (assuming the overlap assumption holds). Moreover, we refer to $w$ as an \textit{observed weight} since it is directly estimable from the data. Section \ref{sec:sens-model-paper} describes our sensitivity model which constrains the worst-case error between $\wast$ and $w$.

\subsection{Related literature} \label{sec:related-lit}
\subsubsection{Causal decomposition analysis} Canonical methods for comparing outcomes across groups include the Kitagawa-Oaxaca-Blinder decomposition which compares fitted values of outcomes for all individuals using a linear model regressed separately on the covariates within each group. \citep{kitagawa1955components,oaxaca1973male, blinder1973wage} VanderWeele and Robinson (2014)\cite{vanderweele2014causal} first proposed intervening on a modifiable exposure when estimating disparities between racial groups, arguing that causal interpretations of race are drawn on the basis of how observed disparities could change if background covariates were equalized across both racial groups. The decomposition into reduced and residual disparity was formalized by Jackson and VanderWeele (2018)\cite{jackson2018decomposition}: They express the two parameters in terms of regression coefficients under the assumption that the outcome and exposure both follow a linear model. Jackson (2021)\cite{jackson2021meaningful} later proposed the weighting-based approach to causal decomposition analysis, which better facilitates the adjustment of a subset of covariates based on notions of equity and the social contract (see Section \ref{sec:allowability-background}) and forms the centerpiece of our sensitivity analysis. Lundberg (2024)\cite{lundberg2024gap} proposed a doubly robust method for decomposing disparities and provides a review on the benefits of disparity decompositions. Recently, Yu and Elwert (2023) \cite{yu2023nonparametric} go beyond the two-part decomposition framework and introduce a four-part, multiply-robust procedure for estimating group disparities and emphasize that, in addition to a baseline disparity, the observed disparity includes components that describe differential selection into treatment, as well as its prevalence and effect. They provide efficient influence functions for each term and demonstrate that they can be estimated using flexible machine learning methods, including double machine learning. \citep{chernozhukov2018double} Ben-Michael et al. (2022) \cite{ben2022estimating} applied both the two-part and four-part decompositions to estimate racial disparities in emergency general surgery, finding in both cases that equalizing the odds of surgery between Black and white patients did not explain observed disparities in adverse outcomes. 

Our framework focuses specifically on \textit{weighted} causal decomposition estimators for which dedicated sensitivity analyses have not previously been proposed. While we do not consider the doubly robust estimators mentioned above further in our subsequent development here, %While the doubly robust estimators mentioned above offer statistical advantages such as precision and variance reduction, they also involve complex functional forms that can be challenging for practitioners to interpret in a sensitivity analysis. Moreover, even if complexity is not of concern, sensitivity analyses for doubly robust estimators are not widely present in traditional observational studies, further complicating its application to causal decompositions. 
 our weighting-based sensitivity framework provides a practical and logical first step towards the development of sensitivity analyses for doubly robust estimators.

\subsubsection{Connection to mediation} Causal decomposition analysis mechanistically follows the same intuition as causal mediation analysis. For instance, the disparity reduction term is akin to the natural indirect effect (NIE) and the residual disparity is analogous to the natural direct effect (NDE).   However, there are several key differences between the two estimands:
\begin{enumerate}
    \item First, identifying the NDE and NIE requires two ignorability assumptions: no treatment-outcome confounding and no mediator-outcome confounding. This is often referred to as \textit{sequential ignorability}. \citep{imai2010identification,ding2024first,Pearl2001DirectAI} Causal decompositions, on the other hand, require only a single conditional ignorability assumption (Assumption \ref{as:condig-paper}) that is weaker than the mediation assumptions. 
    \item Related to the point above, causal mediation analysis assumes both the treatment and mediator are intervenable, forcing multiple ignorability assumptions that are strong and unverifiable. However, this cannot be applied to measuring disparities between groups as it is challenging and often nonsensical to posit potential outcomes with respect to $G$, rendering the NIE and NDE as two uninterpretable and nonexistent estimands. \citep{holland1986statistics} In our application, this would be equivalent to positing suicide rates for a sexual minority youth had we instead fixed their sexual orientation to be heterosexual.
\end{enumerate}
These key differences allow for a causal interpretation of disparity reduction and residual disparity without intervening on group status. We refer the reader to Park et al. (2023) \cite{park2023sensitivity} for a comparison between causal decomposition analysis and various other causal estimands. 

\subsubsection{Sensitivity analysis} While many sensitivity analysis frameworks exist for traditional observational studies, few have been extended to causal decompositions.
\begin{revisiontwo}
   Among the few, Park et al. (2023) \cite{park2023sensitivity} proposes a model-based sensitivity analysis for the causal decomposition framework. In practice, the choice between this framework and ours lies in the estimation strategy: the framework in Park et al. (2023) is only applicable to linear regression outcome models while our framework applies broadly to weighted estimators. In particular, Park et al. (2023) extend the regression-based sensitivity analysis of Cinelli and Hazlett (2020) \cite{cinelli2020making} and re-parameterize the regression coefficients in terms of partial $R^2$ values that describe the unobserved confounder's relationship with treatment and outcome.  
\end{revisiontwo}
A separate paper by Park et al. (2024) \cite{park2024estimation} provides a novel estimation framework and sensitivity analysis to handle multiple exposures, requiring users to specify both an outcome model and a set of weights. This approach is most optimal for multiple mediators but less so in our setting where the number of confounders exceeds the number of mediators. \citep{park2024estimation}

In traditional observational studies, Zhao et al. (2019)\cite{zhao2019sensitivity} adopted the marginal sensitivity model from Tan (2006)\cite{tan2006distributional} to IPW estimation of causal effects. This framework was also utilized by Soriano et al. (2023)\cite{soriano2023interpretable} to handle balancing weights. The RMPW-weighting framework that we adopt was first proposed by Hong (2010) \cite{hong2010ratio} for identifying the NIE/NDE in causal mediation studies and a corresponding sensitivity analysis was developed in Hong et al. (2018)\cite{hong2018weighting} to accommodate violations of the sequential ignorability assumption. While the aforementioned sensitivity analyses were designed for traditional weighted observational studies, it is not trivial to extend them to causal decompositions due to the more complex construction of the RMPW weights in causal decompositions (see Section \ref{sec:comp-zhao}). Our work bridges this gap by adapting weighting-based sensitivity analysis for the novel setting of causal decomposition analysis.

\section{Sensitivity model} \label{sec:sens-model-paper}
\newcommand{\Hlam}{H\paren{\Lambda}}
\newcommand{\wastH}{w^{\ast (h)}}
\newcommand{\muinth}{\mu_{R_0}^{(h)}}

Our sensitivity analysis adopts the marginal sensitivity model (MSM). \citep{tan2006distributional,zhao2019sensitivity,soriano2023interpretable} Traditionally, the MSM establishes worst-case bounds on the odds ratio of the ideal and observed propensity scores in IPW sensitivity analyses. We expand on the IPW approach by considering a generalized form of the MSM: for some $\Lambda \geq 1$, the ideal RMPW weight satisfies
\begin{align}
    \label{eqn:msm-paper}
    \wast \in  \braces{\wast: \Lambda^{-1} \leq \frac{\wast}{w} \leq \Lambda}.
\end{align}
\newcommand{\Lambdaast}{\Lambda^{\ast}}
$\Lambda$ is a sensitivity parameter that constrains the difference between the true and ideal weights. In practice, researchers perform sensitivity analyses under the MSM by recomputing point estimates under increasing levels of $\Lambda$ until the point estimate and/or confidence interval crosses a certain threshold, most commonly zero. We denote this value as $\Lambda^{\ast}$. A study is considered robust to unmeasured confounding if $\Lambda^{\ast}$ is large, indicating that substantial confounding error is required to reverse the results and/or their statistical significance. Conversely, a study is very sensitive to unmeasured confounding if $\Lambda^{\ast}$ is close to 1, suggesting that minimal confounding error could sufficiently overturn the observed findings.

\subsection{Parameter identification under the MSM} \label{sec:msm-paper}
Defining $h  = h(X,U) \coloneqq \log\paren{{\wast}/{w}}$, Zhao et al. (2019)\cite{zhao2019sensitivity} show that the MSM imposes a constraint on the $L_{\infty}$-norm of $h$:
\begin{align}
\label{eqn:shifted-msm-paper}
    H\paren{\Lambda} = \braces{h: \cX \times \R \rightarrow \R : \infnorm{h} \leq \log \Lambda}.
\end{align}
We adhere to this sensitivity model throughout the remainder of the paper. Under this constraint, they introduce a \textit{shifted propensity score} by constraining the log-odds ratio of the ideal and observed propensity scores/weights. Here we construct an analogously modified weight for the RMPW framework. We can express the \textit{shifted ideal weight} as
\begin{align} \label{eqn:shifted-ideal-rmpw-weight}
    w^{(h)} = w\exp\paren{{h(X,U)}},
\end{align}
which is the ideal weight for a particular choice of $h$. The corresponding \textit{shifted estimand} $\muinth$ is represented as
\begin{align}
    \muinth = \frac{\E\paren{w^{(h)} Y \mid G = 1}}{\E\paren{w^{(h)} \mid G = 1}}.
\end{align}
Using the shifted RMPW weights $\hat{w}_i^{(h)} = \hat{w}_i \exp\left({h(X_i, U_i)}\right)$, we can also construct a shifted estimator for $\muint$:
\begin{align}
    \label{eqn:shifted-rmpw-estimator}
    \hat{\mu}_{R_{0}}^{(h)} = \frac{\sum_{i=1}^n G_i \hat{w}_i^{(h)} Y_i}{\sum_{i=1}^n G_i \hat{w}_i^{(h)}}.
\end{align}
Equation \eqref{eqn:shifted-rmpw-estimator} is directly estimable from observed data provided that one specifies the strength of unmeasured confounding through $\Lambda$, which affects $\hat{w}_i^{(h)}$.

The MSM therefore provides a partial identification bound for $\muint$:
\begin{align}
\label{eqn:msm-part-id}
    \inf_{h \in H\paren{\Lambda}} \muinth \leq \muint \leq \sup_{h \in H\paren{\Lambda}} \muinth.
\end{align}
Note that these bounds contain all feasible values of $\muint$ allowed under constraint \eqref{eqn:shifted-msm-paper}, some of which may not be possible to construct due to other constraints on population propensity scores; see Dorn and Guo (2023)\cite{dorn2023sharp} for more discussion. The extrema in Equation \eqref{eqn:msm-part-id} can be efficiently computed using fractional linear programming \citep{zhao2019sensitivity}, which takes the following form:
\begin{align} \label{eqn:extrema-eqn}
    \underset{r_i \in \R^{n_1}}{\text{min/max}} \quad&\hat{\mu}_{R_{0}}^{(h)} = \frac{\sum_{i=1}^n G_i Y_i \brackets{r_i \hat{w}({X_i})}}{\sum_{i=1}^n G_i \brackets{r_i \hat{w}({X_i})}}\\
    \textrm{s.t.} \quad & r_i \in \brackets{\Lambda^{-1}, \Lambda},\nonumber
\end{align}
where $\hat{w}\paren{X_i}$ is the estimated RMPW weight and $n_1 = \sumin G_i$. The decision variables are $r_i = \exp\paren{{h(X_i, U_i)}}$. To assess sensitivity of the disparity reduction term,  $\Lambdaast$ is the critical sensitivity parameter where the disparity reduction becomes zero (i.e., when $\muint = \mu_1$), corresponding to the point where Equation \ref{eqn:extrema-eqn} crosses $\mu_1$. 
In our application, this corresponds to the point where the counterfactual suicidal ideation rate for sexual minorities equals the status quo rate for sexual minorities.

\subsection{Constructing bootstrap confidence intervals} \label{sec:confint-paper}
To account for sampling variability, we follow the sensitivity frameworks in Zhao et al. (2019)\cite{zhao2019sensitivity} and Soriano et al. (2023)\cite{soriano2023interpretable} and use the \textit{percentile bootstrap} to construct confidence intervals for $\hat{\mu}_{R_{0}}^{(h)}$. \smallrevisiontwo{These intervals give asymptotic coverage for both the parameter $\mu_{R_0}$ and its entire partially identified region in Equation \eqref{eqn:msm-part-id}.\cite{imbens2004confidence}}%  Confidence intervals for partially identified parameters such as $\mu_{R_0}$ generally rely on the asymptotic distribution of the boundaries of the partially identified region \citep{imbens2004confidence,huang2024variance}, which is difficult to derive in sensitivity analyses. As a result, the percentile bootstrap approach provides asymptotic point-wise coverage for $\mu_{R_0}$ without requiring explicit knowledge of these distributions.}

We take $B$ bootstrap samples of the full data and re-estimate the weights and the corresponding shifted estimator for each bootstrap sample $b = 1 \dots B$. Let $\hat{\hat{\mu}}^{(h)}_{R_{0}}$ denote the bootstrap distribution of $\muint$ and let $Q_{\alpha}\paren{\cdot}$ denote the $\alpha$-quantile over the $B$ bootstrap replications. Then, for a given $h \in H(\Lambda)$, the percentile bootstrap confidence interval is computed as follows:
\begin{align}
    \label{eqn:boot-ci-h}
    \left[L^{(h)}, U^{(h)}\right] = \left[Q_{\frac{\alpha}{2}}\paren{\hat{\hat{\mu}}^{(h)}_{R_{0}}}, ~Q_{1-\frac{\alpha}{2}}\paren{\hat{\hat{\mu}}^{(h)}_{R_{0}}}\right].
\end{align}

The theorem below states that $\left[L^{(h)}, U^{(h)}\right]$ is an asymptotically valid confidence interval for $\muint^{(h)}$ for a fixed degree of unmeasured confounding, \smallrevisiontwo{and that these intervals can be aggregated to obtain an asymptotically  valid confidence interval for $\muint$.}

\begin{theorem}[Validity of percentile bootstrap] 
\label{thm:valid-pb}
    When $w(X, Z)$ and $\wast(X, Z, U)$ follow a \smallrevisiontwo{parametric} logistic model and each individual weight $\hat{w}_i$ is estimated with logistic regression, then under mild regularity assumptions, \smallrevisiontwo{the following hold:} 

    \begin{align*}
        \underset{n\to\infty}{\lim\sup}\;\mathbb{P}\left(\muint <L^{(h)}\right)\leq\frac{\alpha}{2}
\quad \quad \text{and} \quad \quad
%    \end{align*}
%    and
%    \begin{align*}
        \underset{n\to\infty}{\lim\sup}\;\mathbb{P}\left(\muint >U^{(h)}\right)\leq\frac{\alpha}{2} \quad \quad \text{for every }h \in \Hlam,
    \end{align*}
    \smallrevisiontwo{and $[L,U]$ is an asymptotic confidence interval with $\muint$ with coverage at least $1-\alpha$},
    where $\mathbb{P}$ denotes the joint data-generating distribution of \smallrevisionone{$\paren{G, Z, X, U, Y(1), Y(0)}$,} \smallrevisiontwo{
    and:
\begin{align}
    \label{eqn:boot-ci-cons}
    \left[L, U\right] = \left[Q_{{\alpha}/{2}}\paren{\inf_{h \in H(\Lambda)}{{\hat{\hat{\mu}}^{(h)}_{R_0}}}}, ~Q_{1-{\alpha}/{2}}\paren{\sup_{h \in H(\Lambda)}{\hat{\hat{\mu}}}^{(h)}_{R_0}}\right],
\end{align}
where the extrema inside the quantile functions are computed using the linear program introduced in Equation \eqref{eqn:extrema-eqn}.
}
\end{theorem}
See Section A.3 in the Appendix for the proof, which involves utilizing the $Z$-estimation framework to establish the smoothness of the RMPW weight under mild regularity assumptions, including compact support with the population parameter in the interior of its parameter space and finite fourth-order outcome moments.

Zhao et al. (2019)\cite{zhao2019sensitivity} and Soriano et al. (2023)\cite{soriano2023interpretable} demonstrate that computing a unified confidence interval for $\muinth$ over all possible values of $h$ is computationally expensive, \smallrevisiontwo{but this issue can be mitigated by assigning quantiles of the extrema as confidence bounds, shown in Theorem \ref{thm:valid-pb} Equation \eqref{eqn:boot-ci-cons}.}

\begin{revisionone}
    Since the probability guarantees in Theorem \ref{thm:valid-pb} hold over samples from the joint data-generating process for both the covariates $X$ and the unmeasured confounder $U$, our bootstrap procedure accounts for sampling variability due to both $X$ and $U$ (as do the closely-related procedures of Zhao et al. (2019) \cite{zhao2019sensitivity} and Dorn and Guo (2023) \cite{dorn2023sharp}).  As discussed by Qin and Yang (2022)\cite{qin2022simulation}, sensitivity analyses that ignore the impact of the unmeasured confounder on sampling variability may produce misleading conclusions.%  illustrate how the unmeasured confounder itself affects bootstrap estimates by positing a parametric form for $U$ and repeatedly sampling from its conditional distribution. In the MSM, account for this by sampling from . Our sensitivity analysis inherits the same DGP and bootstrap strategy, and thus the asymptotic results in these frameworks and ours are taken over the entire DGP that includes $U$. 
\end{revisionone}

\subsection{Residual disparity as an equivalence test} \label{sec:equiv-test}
\begin{revisiontwo}
    Standard hypothesis testing for some parameter $\theta$ typically attempts to reject a null of zero effect: 
    \begin{align*}
        H_{0, \text{stnd}}: \theta = 0.
    \end{align*}
    This null cannot be rejected when the confidence interval for $\theta$ crosses zero, leaving open the plausibility of no effect. Scenarios like these can be supplemented with an \textit{equivalence test} which  helps rule out large effect sizes by determining if the effect exceeds a user-specified value. This involves declaring a minimal effect size $\Delta$ and then testing the null hypothesis that the effect is greater than $\Delta$ or less than $-\Delta$:
    \begin{equation*}
       H_{0, \text{equiv}}: \theta > \Delta \text{ or }\theta < -\Delta.
    \end{equation*}
    
    $H_{0, \text{equiv}}$ is equivalent to $\theta \geq \left\vert \Delta \right\vert$. Goeman et al. (2010)\cite{goeman2010three} proposed the ``three-sided test'' which combines $ H_{0,\text{stnd}}$ and $ H_{0, \text{equiv}}$.
    %the equivalence tests and the traditional test of no effect. 
    While a traditional test can only suggest the absence of an effect, the three sided test tells us how small it could be. The individual null hypotheses can be simultaneously tested at level $\alpha$ while maintaining control of the Type I error rate since they are all mutually incompatible -- the risk of falsely rejecting a null hypothesis occurs at most once. \citep{goeman2010three,pimentel2015large} Combining equivalence testing with sensitivity analysis requires repeating the sensitivity analysis with increasing values of $\Lambda$ until the point estimate or confidence interval crosses $\Delta$ or $-\Delta$. The critical parameter $\Lambdaast$ obtained here represents the strength of unmeasured confounding needed to mask an effect of at least $\left\vert \Delta \right\vert$ when the data indicates a small effect is present.

    In causal decompositions, equivalence testing provides a unique perspective on the relationship between the disparity reduction and residual disparity. When the disparity reduction $\mu_1 - \mu_{R_0}$ is small, equivalence testing can be used to assess the null $H_{0, \text{equiv}}$ that $\mu_1 - \mu_{R_0} \geq \left\vert \Delta \right\vert$. For illustration, suppose we set $\Delta = \tau$ (the observed disparity). $H_{0, \text{equiv}}$ then presumes a 100\% reduction in disparity, which is identical to testing for zero residual disparity. Now observe that the critical parameter $\Lambdaast$ obtained from the residual disparity's sensitivity analysis is the degree of unmeasured confounding required to send the residual disparity to zero. Therefore, $\Lambdaast$ can alternatively be interpreted as the degree of unmeasured confounding required to mask a 100\% reduction in disparity -- performing sensitivity analysis of the residual disparity mirrors equivalence testing for the disparity reduction. In practice, researchers can specify any null percentage of disparity reduction $100\eta\% ~\forall \eta \in (0,1]$ and perform a three-sided test. We demonstrate the three-sided test for the ABCD study in Appendix \ref{sec:app-robustness-Z}, where we find a small and insignificant disparity reduction effect.
\end{revisiontwo}

\subsection{Comparison to MSM for Inverse Propensity Score Weighting} \label{sec:comp-zhao}
\newcommand{\wipw}{w_{\scriptscriptstyle{\mathrm{IPW}}}}
\newcommand{\wipwh}{w^{(h)}_{\scriptscriptstyle{\mathrm{IPW}}}}
\newcommand{\wipwast}{w^{\ast}_{\scriptscriptstyle{\mathrm{IPW}}}}

\begin{revisiontwo}
    The MSM for IPW\citep{tan2006distributional,zhao2019sensitivity,soriano2023interpretable}, while equivalent to the MSM introduced earlier in this section, is illustrated in a different way. Broadly speaking, the IPW-MSM places bounds on the \textit{odds ratio} of the propensity scores whereas our version bounds the weights directly. Here we demonstrate how the IPW-MSM is a specific case of our generalized MSM.

    Recall from Equation \eqref{eqn:msm-paper} that our MSM satisfies 
    \begin{align*}
        {\wast: \Lambda^{-1} \leq \frac{\wast}{w} \leq \Lambda}. 
    \end{align*}
    Now let $e(X)$ and $\exu$ denote the observed and ideal propensity scores, respectively. The IPW MSM \citep{tan2006distributional,zhao2019sensitivity,soriano2023interpretable} is characterized as
    \begin{align*}
        {\exu: \Lambda^{-1} \leq \mathrm{OR}\braces{\ex, \exu} \leq \Lambda},
    \end{align*}
    where $\mathrm{OR}\braces{\cdot, \cdot}$ is the odds ratio. The key is that $\mathrm{OR}\braces{\ex, \exu}$ is equivalent to a monotone transformation of the inverse propensity score weight:
    \begin{align*}
        \mathrm{OR}\braces{\ex, \exu} = \frac{\wipwast - 1}{\wipw - 1},
    \end{align*}
    where $\wipw =1/ \ex$ and $\wipwast = 1 / \exu$. 

    This monotone transformation also causes the shifted weights to have different expressions:
    \begin{align*} 
        w^{(h)} &= w\exp\brackets{{h(X,U)}}\qquad\text{(general MSM)}\\
        \wipwh - 1 &= {\paren{\wipw - 1} \exp\brackets{{h(X,U)}}}\qquad\text{(IPW-MSM)}.
    \end{align*}
    Therefore, we can express the IPW-MSM in terms of our generalized MSM by setting $w = \wipw - 1$ and $\wast = \wipwast - 1$. Because $\wipw$ is simply the reciprocal of the propensity score, the odds ratio approach has a more straightforward interpretation. By comparison, the RMPW weight has a more complex structure that precludes interpretation using the propensity scores, making weights a more practical way to reason about unmeasured confounding. Unifying both frameworks demonstrates that the IPW-MSM can also be parameterized as a ratio of weights despite being illustrated as an odds ratio of propensity scores. As a result, our general MSM formulation achieves the same inferential conclusions from Zhao et al. (2019)\cite{zhao2019sensitivity} and Soriano et al. (2023).\cite{soriano2023interpretable}
\end{revisiontwo}

\subsection{Sensitivity analysis on the ABCD Study} \label{sec:abcd-initial}
We illustrate our proposed sensitivity analysis on the ABCD Study by examining the effect of parental support on suicidal ideation in sexual minority youth. We estimate the observed disparity, disparity reduction, and residual disparity. We also include two critical values of $\Lambda$ from the MSM: $\Lambda^{\ast}_{0.05}$ corresponding to the value where the 95\% confidence interval crosses zero and $\Lambda^{\ast}$ where the point estimate bound crosses zero. The results can be found in Table \ref{tbl:abcd-initial-aoas}. Note that $\Lambda^{\ast}_{0.05}$ is computed using the percentile bootstrap procedure, which is subject to Monte Carlo error. Standard errors for the disparity reduction and residual disparity were computed using the bootstrap as suggested in Jackson (2021).\cite{jackson2021meaningful} Propensity score models for $e_1$ and $e_{0a}$ were constructed using logistic regression with covariates and their two-way interactions. Following the allowability rubric \citep{jackson2021meaningful}, we designate allowable covariates as age and sex and non-allowable covariates as sibling order and number of siblings, income, family conflict, peer victimization, school safety, neighborhood safety, neighborhood area deprivation index (ADI), and state-level structural stigma against sexual minorities.

\begin{table}[ht] 
\centering 
 \begin{tabular}{lcccc} 
 \hline 
  \textbf{Parameter} & \textbf{Estimate (SD)} & \textbf{95\% CI} & $\Lambda^{\ast}_{0.05}$ & $\Lambda^{\ast}$ \\ \hline
 Observed Disparity ($\mu_1 - \mu_0$) & 0.251 (0.019) & [0.214, 0.288] & -- & --\\ 
 Disparity Reduction ($\mu_1 - \muint$) & 0.04 (0.016) & [0.006, 0.069] & 1.02 & 1.09\\ 
 Residual Disparity ($\muint - \mu_0$) & 0.211 (0.02) & [0.168, 0.257] & 1.53 & 1.68\\ \hline\hline 
 \end{tabular} 
 \caption{Observed disparity, disparity reduction, residual disparity, and critical sensitivity parameters and 95\% confidence intervals for the ABCD Study. $\Lambda^{\ast}_{0.05}$ corresponds to the critical parameter where the \textit{confidence interval} crosses 0, and $\Lambda^{\ast}$ corresponds to the critical parameter where the \textit{point estimate} crosses 0. There are no critical sensitivity parameters for the observed disparity since it is not a causal estimand.} 
\label{tbl:abcd-initial-aoas} 
\end{table} 

The observed disparity is the difference in proportions of sexual minority and heterosexual youth who had suicidal ideation. A simple difference in means shows that the suicidal ideation rate is 0.251 greater than heterosexuals ($\mu_1 = 0.451$, $\mu_0 = 0.20$). After controlling for factors such as age and sex in both groups, the base disparity remains high (0.23). This disparity is concordant with previous studies of suicidality, particularly the meta-analysis discussed in Section \ref{sec:introduction} which reported $\mu_1 = 0.28$ and $\mu_0 = 0.12$. \citep{marshal2011suicidality} The lower group rates and base disparity could be due to its inclusion of all forms of suicidality, including suicide attempts.

If the distribution of parental support for sexual minorities followed that of heterosexuals, the suicidal ideation rate could be reduced by roughly 0.04 ($\muint = 0.411$, 16\% of the overall disparity), providing a prima facie indication that parental support may not reduce the disparity by a large fraction. An analogous statement can be made for the residual disparity. In Sections \ref{sec:amplification} and \ref{sec:application}, we introduce and demonstrate our reformulated sensitivity analysis, allowing us to visualize, interpret and calibrate these results with respect to observed confounders. 

\section{Amplification of the MSM} \label{sec:amplification}
% Amplification Section
\begin{revisionone}
    Recall that a variable must be associated with both the intervention and the outcome to be considered a confounder. The MSM only bounds an unmeasured confounder's relationship with the intervention, as shown in Equation \eqref{eqn:msm-paper}. This implicitly assumes a worst-case bound on the confounder-outcome relationship which may lead to impractical and overly conservative results for researchers trying to identify realistic threats to their study. Furthermore, reasoning about the plausibility of an unmeasured confounder with strength $\Lambdaast$ can be challenging, as this parameter reflects a distributional difference that may not be straightforward to interpret. 
    
    To address these challenges, we introduce a two-parameter representation of the MSM, offering a deeper and more intuitive perspective of unmeasured confounding by directly modeling the confounder-outcome relationship. More specifically, our two-parameter representation is an \textit{amplification},\cite{rosenbaum2009amplification,soriano2023interpretable} or a mapping between a one-dimensional sensitivity analysis and an equivalent multi-dimensional sensitivity analysis. Researchers may also prefer to characterize the amplification as a \textit{secondary sensitivity analysis} that is more interpretable than the MSM while still enjoying its statistical guarantees. 
    Under a linear model assumption, the MSM parameter ${\Lambda}$ can be re-parameterized as two terms: one describing its relationship with the outcome and the other describing its imbalance. This approach decomposes the intervention and outcome mechanisms of unmeasured confounding and offers increased interpretability by grounding the sensitivity parameters within a familiar linear regression framework. The use of both tools allows researchers to compute a one-dimensional sensitivity analysis with the MSM and interpret the results using its amplified parameters. Soriano et al. (2023)\cite{soriano2023interpretable} introduced a similarly structured amplification for average treatment effect (ATE) estimation. 
\end{revisionone}

\subsection{Worst-case bias as a product of two terms} \label{sec:amp-deriv-paper}
Assume without loss of generality that $U$ is centered and scaled to have mean 0 and variance 1. To guide interpretability, we posit the following working models for the expected potential outcomes, conditional on $X$ and $U$:
\begin{align}
    \E\brackets{Y(1) \mid G = 1, X, U} &= \beta_z + f(X) + \beta_u U \label{eqn:model-y1} \\
    \E\brackets{Y(0) \mid G = 1, X, U} &= f(X) + \beta_u U \label{eqn:model-y0}.
\end{align}
Thus, when modeling the observed outcome $Y = Z Y(1) + (1-Z) Y(0)$, we have
\begin{align*}
    \E\brackets{Y \mid G = 1, X, U} = \beta_z Z + f(X) + \beta_u U.
\end{align*}
When the distribution of treatment is equalized between groups, the true conditional expectation of the counterfactual outcome $Y(int)$ for group $G=1$ is:
\begin{align}
\label{eqn:true-Yint}
    \E\brackets{Y(int) \mid G = 1, X, U} &= P(Z=1 \mid G = 0, X, U)~\E\brackets{Y(1) \mid G = 1, X, U}+ \nonumber\\
    &\quad P(Z=0 \mid G = 0, X, U)~\E\brackets{Y(0) \mid G = 1, X, U}\nonumber\\
    &= e_0^{\ast}~\E\brackets{Y(1) \mid G = 1, X, U} + (1-e_0^{\ast})~\E\brackets{Y(0) \mid G = 1, X, U}
\end{align}
When ignorability holds, the counterfactual outcome can be identified using the RMPW estimand $\E\brackets{wY \mid G = 1}$, which reweights the observed outcome by the RMPW weight defined in Section \ref{sec:setup}. In general, the true conditional form of the RMPW estimand is expressed as
\begin{align}
\label{eqn:ignor-Yint}
    \E\brackets{wY \mid G = 1, X, U} &= \E\brackets{w\paren{ZY(1) + (1-Z) Y(0)} \mid G = 1, X, U}\nonumber\\
    &= \frac{e_0}{e_1} \E\brackets{Z Y(1) \mid G = 1, X, U} + \frac{1-e_0}{1-e_1}\E\brackets{(1-Z) Y(0) \mid G = 1, X, U}.
\end{align}
Taking the expectation of the difference between Equations \eqref{eqn:true-Yint} and \eqref{eqn:ignor-Yint} yields the bias of the RMPW estimand with respect to the true functional form of $Y(int)$, which we introduce in the following theorem:  

\begin{theorem}[Ignorability bias decomposition] 
\label{thm:amp-bias-decomp}
    Suppose the working models posited in Equations \eqref{eqn:model-y1} and \eqref{eqn:model-y0} hold for $\E\brackets{Y(1) \mid G = 1, X, U}$ and $\E\brackets{Y(0) \mid G = 1, X, U}$, respectively. Suppose also that $U$ is non-allowable. Then, the population-level bias between the true counterfactual outcome in group $G=1$ and the RMPW estimator in group $G=1$ can be written as
    \begin{align}
        \mathrm{Bias}\left(Y(int), ~wY\right) &= \E\brackets{Y(int) \mid G=1} - \E\brackets{wY \mid G = 1} \nonumber\\
        &= \beta_u \delta_u,
    \end{align}
    where
    \begin{align}
    \label{eqn:delta-u}
        \delta_u = \E\brackets{\frac{e_0 - e_1}{1 - e_1} \paren{U - \frac{ZU}{e_1}} \mid G = 1}.
    \end{align}
\end{theorem}

\begin{corollary}
\label{cor:part-id-bias}
    Under the same assumptions as Theorem \ref{thm:amp-bias-decomp}, for a given value of $\Lambda$, the bias can be lower and upper-bounded as
    \begin{align*}
        \inf_{h \in H({\Lambda})} {\mu}_{R_{0}}^{(h)} - \E\brackets{w Y \mid G = 1} \leq {\beta_u}{\delta_u} \leq \sup_{h \in H({\Lambda})} {\mu}_{R_{0}}^{(h)} - \E\brackets{w Y \mid G = 1}.
    \end{align*}
\end{corollary}
Under one additional assumption, $G \indep U \mid X^A$ \citep{park2023sensitivity}, Theorem \ref{thm:amp-bias-decomp} also holds for unmeasured \textit{allowable} confounders or when the allowability framework is not used. This assumption states that sexual minority status is independent of the unobserved confounder within levels of allowable covariates. However, since a majority of the covariates we consider are non-allowable (see Section \ref{sec:abcd-initial} for more discussion), it is more plausible that an unmeasured covariate would be non-allowable. The proofs can be found in Section A.4 in the Appendix.

$\beta_u$ is a scalar regression coefficient that represents the change in outcome corresponding to a one standard deviation increase in $U$. $\delta_u$ is the expected difference in $U$ between the $G=1$ group as a whole and those who receive the intervention in the $G=1$ group, multiplied by a scaling factor. This factor is entirely determined by the observed group-level propensity scores and is less than 1 if $e_0 > e_1$ and equal to zero if $e_0 = e_1$. This implies that $U$ would be less imbalanced for an individual if their fitted probability of superior parental support as a heterosexual is higher than as a sexual minority. The imbalance and bias are both zero if the group propensity scores $e_0$ and $e_1$ are equal, indicating that the policy defining the intervention is the same across both groups. In our application, $\delta_u$ measures the imbalance of $U$ between the overall sexual minority population and the sexual minority population that receives superior parental support.
\begin{revisiontwo}
    Consider a simple example where $e_1 = 0.4$ and $e_0 = 2e_1$ for all individuals, where exactly half have superior parental support ($Z=1$) and the other half do not ($Z=0$). Moreover, assume $U$ is a binary variable that is equal to 1 if $Z=0$ and 0 otherwise (an inverse proxy for $Z$). Using Equation \ref{eqn:delta-u} in Theorem \ref{thm:amp-bias-decomp}, it follows that $\delta_u = 1/3$ in this scenario. 
    If we instead let $e_1 = 0.2$ and keep everything else the same, then $\delta_u = 1/8$. 
\end{revisiontwo}
 
% \begin{revisiontwo}
%     One may also consider $e_0$ as a constant multiplier of $e_1$ such that $e_0 = c\cdot e_1$. This approach restricts the amplification to the $G=1$ group, making it advantageous if researchers are comfortable reasoning about $e_0$ in terms of $e_1$. Under this simplification, the imbalance term $\delta_u$ reduces to
%     \begin{align*}
%         \delta_u = (c-1)~{\E\brackets{\frac{e_1}{1-e_1} \paren{U - \frac{ZU}{e_1}} \mid G = 1}},
%     \end{align*}
%     which only depends on quantities in the $G=1$ group. This refinement may be useful if covariates in the $G=0$ group contain a lot of missing values. However, the outcomes $Y$ for both groups are required in order to compute the residual disparity term and $\tau$. Pursuing this approach requires sufficient domain knowledge to justify the constant $c$ in order to ensure the parameters are realistic.   
% \end{revisiontwo}

Corollary \ref{cor:part-id-bias} establishes the connection between $(\beta_u, \delta_u)$ and the MSM by leveraging the fact that $\muint$ is partially identified for any $h \in H({\Lambda})$:
\begin{align*}
    \inf_{h \in H({\Lambda})} {\mu}_{R_{0}}^{(h)} \leq \mu_{R_{0}} \leq \sup_{h \in H({\Lambda})} {\mu}_{R_{0}}^{(h)}.
\end{align*}
\begin{revisiontwo}
    As mentioned previously, our amplification can also be viewed as a secondary sensitivity model with different parameters $(\beta_u, \delta_u)$ that shares the same partial identification bounds as the MSM. These bounds can be used to place an upper bound on the maximum absolute bias from Theorem \ref{thm:amp-bias-decomp}:
\end{revisiontwo}
% Therefore, we can upper bound the maximum bias as
\begin{align*}
    \left|\beta_u \cdot \delta_u\right| \leq \max \left\{ \left| \inf_{h \in H(\Tilde{\Lambda})} {\mu}_{R_{0}}^{(h)} - {\mu}_{R_{0}} \right|, \left| \sup_{h \in H(\Tilde{\Lambda})} {\mu}_{R_{0}}^{(h)} - {\mu}_{R_{0}} \right| \right\}.
\end{align*}
Finite-sample bias estimates can be computed by replacing the population-level estimands with their finite-sample analogues. This establishes a closed-form upper bound of the absolute bias for a given sensitivity model using the linear program discussed in Section \ref{sec:msm-paper}. \smallrevisiontwo{Using this regression-based characterization of a weighted sensitivity analysis allows researchers to interpret results from the MSM in a more familiar setting while enjoying its statistical and inferential guarantees.}

Observe that the functional form of the observed covariates $f(X)$ and the linear coefficient of $U$ need not reflect their true relationship with $Y(int)$; rather, the role of $\beta_u$ is to quantify the relationship between $U$ and $Y(int)$ that contributes to the bias rather than a correctly-specified model. \citep{cinelli2020making,soriano2023interpretable} The working models introduced in Equations \eqref{eqn:model-y1} and \eqref{eqn:model-y0} improve interpretability but assume a functional form for the expected potential outcome. Researchers who prefer to avoid these assumptions may utilize the MSM described in Section \ref{sec:sens-model-paper} which is fully nonparametric. Our sensitivity framework provides users with flexibility to decide whether the amplification is advantageous and to use it accordingly.

\subsection{A tool for interpretation and calibration}
\begin{revisionone}
    Imposing an outcome model and simplifying interpretation makes our amplification well-suited and accessible for applied researchers. Since causal decomposition analysis is rooted in practical applications of health disparity research, this reparametrization of the MSM also allows researchers to calibrate their results in the context of observed covariates, ensuring their results are robust and directly applicable to the study at hand.
\end{revisionone}

\smallrevisiontwo{Building on Soriano et al. (2023)\cite{soriano2023interpretable} and Imbens (2003) \cite{imbens2003sensitivity}, we employ a similar calibration procedure by treating each covariate as if it were omitted and computing its $(\beta_u, \delta_u)$ pair. This approach} offers a broad sense of plausible parameter values of $U$ based on observed covariates. %\smallrevisiontwo{Cinelli and Hazlett (2020)\cite{cinelli2020making} propose a formal calibration (benchmarking) method in the context of outcome regression sensitivity analysis and we relegate this as an area of future work for the marginal sensitivity model.}

These calibration procedures can be visualized using two-dimensional \textit{bias contour plots}: A given bias value can be parameterized by a grid of various $(\beta_u, \delta_u)$ pairs which trace out a single contour. For instance, a $(\beta_u, \delta_u)$ pair of $(1,1)$ and $(2, 0.5)$ both correspond to the same bias of 1 but are distributed differently across impact and imbalance: A strongly prognostic covariate ($\beta_u$) need not be very imbalanced ($\delta_u$) to induce the same amount of bias, and vice versa. We are interested in the \textit{critical bias} where the point estimates or bootstrap confidence interval crosses 0, corresponding to $\Lambdaast$ in the MSM. Values of $\beta_u$ and $\delta_u$ that result in greater bias indicate the respective impact and imbalance necessary to nullify an observed result. We refer to these confounders as \textit{killer confounders} since they correspond to bias values that can substantively alter a research conclusion. \citep{hartman_huang_survey_sensitivity,huang2024sensitivity,huang2024overlap} 
Calibrated $(\beta_u, \delta_u)$ points from observed covariates provide a sense of how unmeasured confounders contribute to the bias if they behaved like observed covariates and whether they could be killer confounders. While these procedures and aids cannot reduce a study's unmeasured confounding, they provide a transparent platform for researchers to reason about the strength and/or plausibility of unmeasured confounders. \citep{cinelli2020making}
Moreover, Cinelli and Hazlett (2020)\cite{cinelli2020making} propose a formal calibration (benchmarking) method in the context of regression-based sensitivity analysis and we relegate this as an area of future work for the marginal sensitivity model.

\section{Application: Suicidal ideation in sexual minority youth} \label{sec:application}
\begin{figure}[ht]
    \centering
    \includegraphics[width=0.75\linewidth]{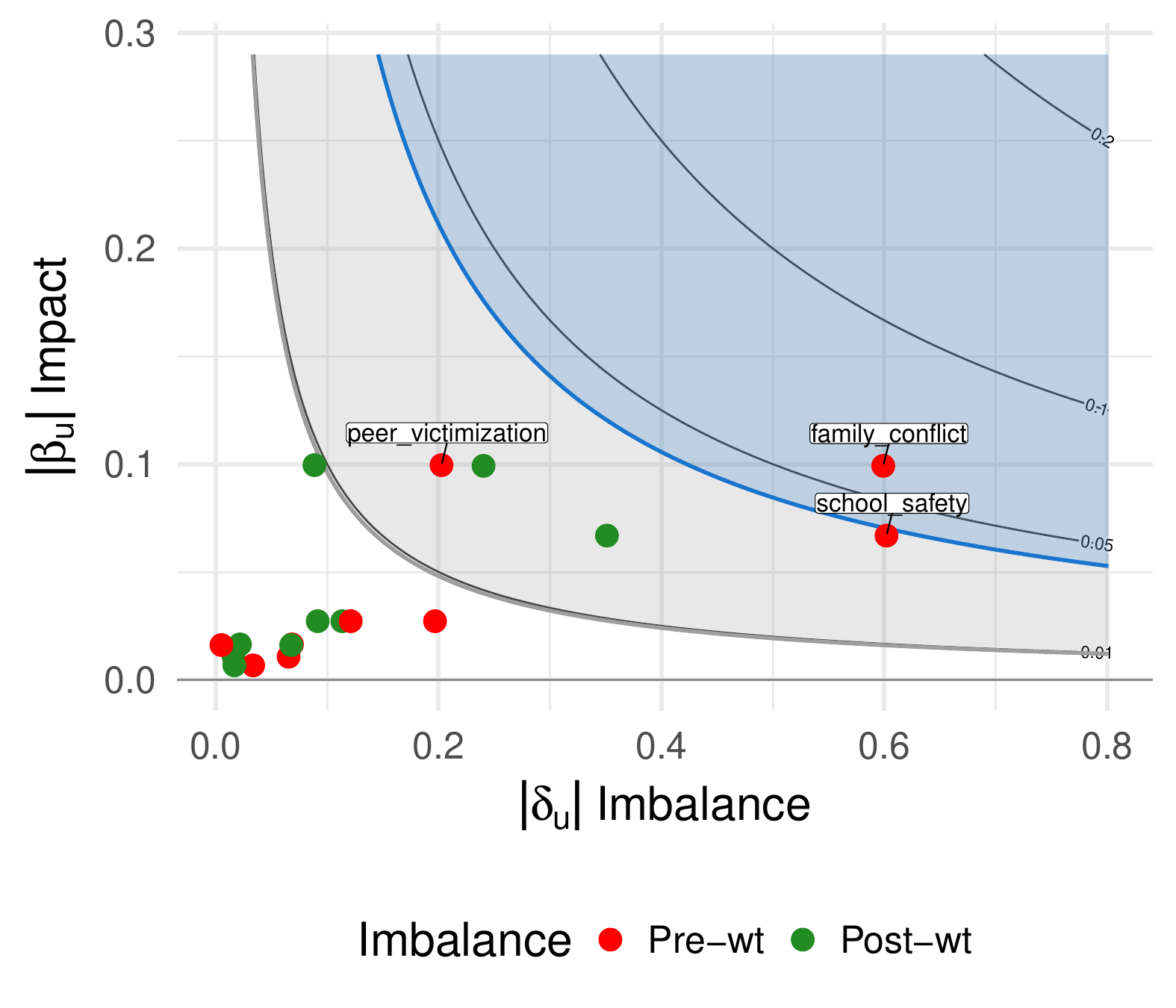}
    \caption{Bias contour plot of {\textbf{disparity reduction}} for the ABCD study. We plot $\delta_u$ on the $x$-axis and $\beta_u$ on the $y$-axis. Red points correspond to pre-weighting imbalance and green points correspond to post-weighting imbalance. The gray curve and region represents the bias corresponding to $\Lambda^{\ast} = 1.02$ where the result is no longer statistically significant at level $\alpha = 0.05$ but the point estimate has not yet reached zero. The blue curve represents the bias corresponding to $\Lambda^{\ast} = 1.09$ where the point estimate crosses zero. The blue shaded region corresponds to the killer confounder region where the bias is large enough to erode the point estimate. Labels are provided for the top three covariates with greatest bias ($\left\vert \beta_u \delta_u\right\vert$).}
    \label{fig:reduction-abcd-contour}
\end{figure}

\begin{figure}[ht]
    \centering
    \includegraphics[width=0.75\linewidth]{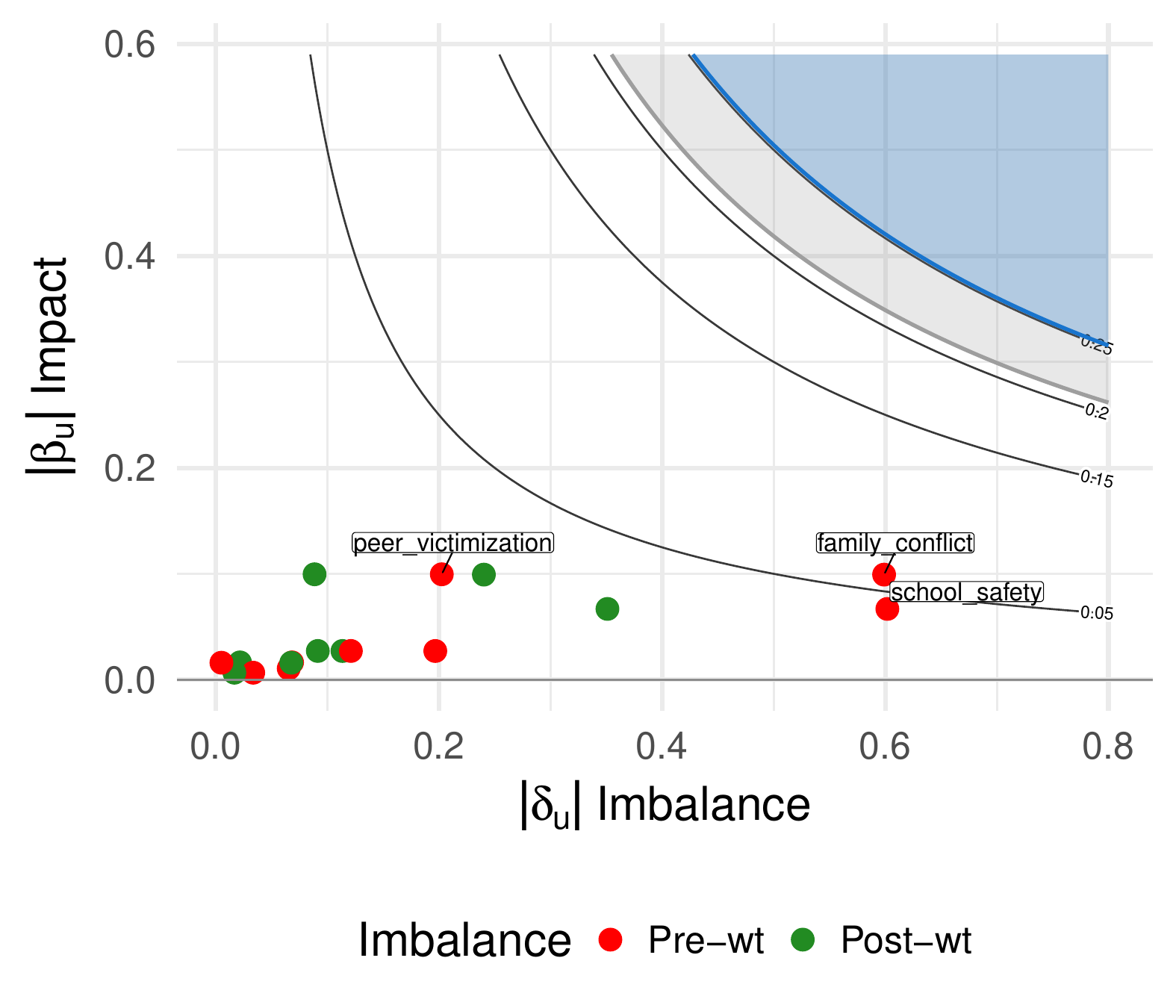}
    \caption{Bias contour plot of {\textbf{residual disparity}} for the ABCD study which serves as an equivalence test for the disparity reduction. We plot $\delta_u$ on the $x$-axis and $\beta_u$ on the $y$-axis. Red points correspond to pre-weighting imbalance and green points correspond to post-weighting imbalance. The gray curve and region represents the bias corresponding to $\Lambda^{\ast} = 1.53$ where the result is no longer statistically significant at level $\alpha = 0.05$ but the point estimate has not yet reached zero. The blue curve represents the bias corresponding to $\Lambda^{\ast} = 1.68$ where the point estimate crosses zero. The blue shaded region corresponds to the killer confounder region where the bias is large enough to mask a 100\% disparity reduction. Labels are provided for the top three covariates with greatest bias ($\left\vert \beta_u \delta_u\right\vert$).}
    \label{fig:residual-abcd-contour}
\end{figure}

Building on the initial sensitivity analysis in Section \ref{sec:abcd-initial}, we now return to the ABCD Study to demonstrate our amplification. Using the values of $\Lambdaast$ introduced in Table \ref{tbl:abcd-initial-aoas} of Section \ref{sec:abcd-initial}, we compute the maximum bias for the disparity reduction (0.042 for $\Lambdaast = 1.09$) and residual disparity (0.252 for $\Lambdaast = 1.68$) and trace out the corresponding bias contour curve, shown in Figures \ref{fig:reduction-abcd-contour} and \ref{fig:residual-abcd-contour}. The killer confounder region (in blue) contains all values of $(\beta_u, \delta_u)$ that result in a greater bias than the bias for $\Lambdaast$. Note that our plotted killer confounder regions correspond to the strength of unmeasured confounding required to reduce the point estimate bound to zero. \smallrevisionone{The region in gray contains $(\beta_u, \delta_u)$ pairs such that the estimated effect is no longer statistically significant at level $\alpha = 0.05$ but not yet zero.}  

Figure \ref{fig:reduction-abcd-contour} displays the bias contour plot for the disparity reduction term. The horizontal axis depicts the absolute imbalance ($\left\vert \delta_u \right\vert$) which is the difference in standardized covariates between the overall sexual minority ($G = 1$) and treated sexual minority ($ZG = 1$) groups. The vertical axis plots the absolute multiple regression coefficient ($\left\vert \beta_u \right\vert$) of each standardized covariate if it was the sole unmeasured confounder (impact). 

\begin{revisiontwo}
    The colored points represent different versions of imbalance that are shifted along the horizontal axis. Red points represent the pre-weighting imbalance in $U$: $\mathbb{E}\left[U - ZU \mid G = 1\right]$. This term is a modified version of $\delta_u$ that sets the scaling factor of propensity scores in Equation \ref{eqn:delta-u} equal to 1 and does not reweight $ZU$ by $e_1$. By construction, this term is equal to $\mathbb{E}\left[(1-Z)U \mid G = 1\right]$ which simplifies to $-\mathbb{E}\left[ZU \mid G = 1\right]$ because $U$ has mean 0. Therefore, the pre-weighting imbalance can be interpreted as the degree to which $U$ skews in the control group before weighting is applied, equivalent to the negative of this value for the treated group. Therefore, large values of $\left\vert\mathbb{E}\left[ZU \mid G = 1\right]\right\vert$ reflect greater intrinsic differences in $U$ between treated and untreated sexual minorities. This metric provides a simplified way to assess imbalance prior to weighting. Green points represent post-weighting imbalance which is expressed as $\delta_u$ from Theorem \ref{thm:amp-bias-decomp}.
\end{revisiontwo}

Like Soriano et al. (2023)\cite{soriano2023interpretable}, green points provide more optimistic imbalance estimates for an unmeasured confounder since they are computed with respect to observed covariates which are easier to balance. Conversely, the red points represent pre-weighting imbalance and serve as a heuristic approximation for how imbalanced an unmeasured confounder may actually be.

While the plot indicates strong plausibility of unmeasured confounding, this statement must be considered with respect to the observed covariates: Covariates such as family conflict and school safety are more imbalanced (higher $\delta_u$) compared with other non-allowable covariates, before and after weighting, which drives the sensitivity of the results. In particular, these two covariates exhibit the greatest pre-weighting imbalance and an omitted confounder with the same $\beta_u$ need only have a post-weighting imbalance two-thirds as large as the observed pre-weighting imbalance of family conflict ($\delta_u \approx 0.4$) to substantially alter the disparity reduction. Peer victimization, although similarly predictive of suicidal ideation, is less imbalanced than school safety or family conflict. This suggests that sexual minorities, regardless of parental support, experience more similar degrees of peer victimization compared to the safety of schools they attend or their family conflict. Given the substantial influence of family conflict on parental support, our sensitivity analysis prompts researchers to consider whether other equally or more influential factors affecting parental support and/or suicidal ideation might have been omitted from the study.

\begin{revisiontwo}
    A possible unmeasured confounder that could drive the relationship between parental support and suicidal ideation is the expression of a hereditary genetic factor that predisposes parents to poor mental health, precluding them from properly supporting their children. The gene is passed onto the child which places them at a higher risk of suicidal ideation. We would expect a genetic factor like this to exhibit a large degree of imbalance between kids with high and low parental support, perhaps as imbalanced as family conflict. Therefore, a one standard deviation increase in gene expression need only increase the probability of suicidal ideation by roughly $2.5$\% ($\beta_u \approx 0.025$) to reverse statistical significance or by roughly $7.5$\% ($\beta_u \approx 0.075$) to completely nullify any disparity reduction. 
\end{revisiontwo}

Figure \ref{fig:residual-abcd-contour} shows the bias contour plot for the residual disparity term, which also serves as an equivalence test for the disparity reduction as described in Section \ref{sec:equiv-test}. Since $\Lambdaast$ for the disparity reduction is less than $\Lambdaast$ for the residual disparity, we conclude that the former is more sensitive to unmeasured confounding, and an unmeasured confounder strong enough to mask a large portion of the disparity reduction in our equivalence test would already be able to nullify the observed disparity reduction and reverse its sign. In order to reason about masking a complete disparity reduction, one must posit an unmeasured confounder with six times as much bias (0.252) as an unmeasured confounder required to nullify the observed disparity reduction (0.042). These covariates would need to exhibit the same amount of imbalance as family conflict or school safety and over four times as much impact. Because such confounders seem unlikely to exist, our analysis suggests that intervening on parental support will result in a modest reduction in disparity at best.

One limitation of our data analysis is the potential for inappropriate temporal ordering of variables or reverse causation; we assume that treatments are measured prior to outcomes but we define our outcome using all four timepoints in the ABCD data. This leaves open the possibility that observed associations between parental support and suicidal ideation may be at least partly due to effects of the latter on the former. As a robustness check, we repeat the sensitivity analysis after excluding youth who had suicidal ideation at baseline, ensuring that outcomes follow after our initial measurement of parental support. The results agree qualitatively with the ones presented in this section, suggesting that any reverse causation is limited in its impact. See Section B in the Appendix for more details.

We emphasize that our sensitivity analysis framework is meant to equip researchers with tools to help reason about unmeasured confounding in their study and is {not} intended to serve as the sole basis for determining whether unmeasured confounding exists or its degree. Researchers must incorporate their own expertise to determine the plausibility of unmeasured confounders as strong as or stronger than the observed ones and whether they are able to alter a study's result. Exercises like that of the preceding paragraphs in this section shift the discussion from the general threat of unmeasured confounding to a principled argument of when that threat is problematic and when it may be safely ignored. \citep{cinelli2020making}

\section{Discussion} \label{sec:discussion}
Identifying target interventions for disparity mitigation is an important step towards a more equitable society where health and well-being are not substantially lower among underrepresented groups. Causal decomposition analysis evaluates hypothetical target interventions from observational data by estimating the disparity eliminated after counterfactually equalizing its access amongst groups. These analyses also suffer from unmeasured confounding, and our paper develops a sensitivity analysis framework for weighted causal decomposition estimators using the marginal sensitivity model. To enhance interpretability, we offer a two-parameter amplification that allows researchers to visualize their study's confounding mechanism and cogently reason about unmeasured confounders in terms of their impact and imbalance. We demonstrate the utility of our sensitivity analysis by scrutinizing parental support as a potential mitigating factor for suicide risk in sexual minority youth. Our application highlights that any reduction in disparity is modest at best and suggests comparing these results to the effects of other target interventions.

We suggest several avenues for future work. Since our analysis did not suggest parental support as a strong mechanism for disparity reduction, further screening studies should be conducted to identify interventions that demonstrate greater efficacy. Once promising candidate interventions are identified, researchers might proceed by designing optimal strategies for deploying the intervention and conducting randomized trials to confirm their hypotheses. Since target interventions will likely be indirect and take the form of encouragement designs, issues of compliance will likely play a central role in the  design and analysis of such trials.

On the methodological side, the marginal sensitivity model provides conservative confidence interval estimates. In the IPW setting, these bounds were sharpened using quantile balancing \citep{dorn2023sharp} which constrains balance on outcome quantiles. Providing sharper bounds to the bootstrap intervals would allow for more informative inference about the change in disparity. Regarding the renfinement of bounds, Huang and Pimentel (2024)\cite{huang2024variance} propose the \textit{variance-based sensitivity model} which parameterizes the residual variation between  $w^{\ast}$ and $w$. Such models rely on an assumption of the weights that $\mathbb{E}\brackets{w^{\ast} \mid G = 1} = w$, which may not always be the case in the RMPW setting. Exploring variance-based sensitivity approaches to causal decomposition analysis may allow for stronger inferential conclusions in a different context and could be compared against the MSM for a more robust argument for bolstering one's decomposition analysis.

The RMPW weights are computed using a plugin approach, with propensity scores estimated using logistic regression. Even when models for both $e_0$ and $e_1$ are well-specified, such weights are only guaranteed to balance covariates approximately and in large samples. Plugin weights may also exhibit high variance. Balancing weights improve finite-sample performance by solving for weights that satisfy balance constraints while minimizing variance. \citep{hainmueller2012entropy,ben2021balancing} These methods may be understood as regularized approaches to propensity score modeling. \citep{zubizarreta2015stable,wang2020minimal} Designing a balancing weights estimation scheme for causal decompositions along with a corresponding sensitivity framework would allow for weights that satisfy balance constraints upfront rather than forcing researchers to perform post-hoc balance checks.

Regarding the allowability framework, future work could consider alternative ways to give insight into multiple confounding mechanisms.
%since our amplification requires an extra assumption for allowable unmeasured confounders. 
Jackson (2021)\cite{jackson2021meaningful} further partitions allowable covariates into those that pertain specifically to the target intervention and outcome. However, this results in slightly more complicated weights and estimators. Following the sensitivity analysis from Park et al. (2023)\cite{park2023sensitivity} and the causal decomposition framework from Yu and Elwert (2023)\cite{yu2023nonparametric}, we do not distinguish between different types of allowable covariates and group them together instead. A next step could carefully consider what robustness to different kinds of unmeasured confounders would look like.

Finally, our two-parameter amplification, while more interpretable than the single-parameter MSM, suffers from issues of scale. Cinelli and Hazlett (2020)\cite{cinelli2020making} discuss how the omitted-variable bias (OVB) framework in regression is easiest to understand for binary confounders but more difficult when the scale of a confounder is not easily measurable. While standardization mitigates the correctness of the parameters, it may muddle interpretation of more abstract confounders such as discrimination intensity. \citep{park2023sensitivity} Cinelli and Hazlett (2020)\cite{cinelli2020making} show how the traditional OVB parameters can be re-expressed as scale free partial $R^2$ terms, and this approach was extended in linear causal decomposition models\citep{park2023sensitivity} as discussed in Section \ref{sec:related-lit}. However, the impact/imbalance parameters in our weighting-based amplification have slightly different interpretations from the OVB setting. Re-expressing the implied model in our amplification in terms of scale free parameters would provide a crucial first step in unifying the weighting and regression frameworks in the causal decomposition setting which is a presently active area of research in traditional causal inference. \citep{chattopadhyay2023implied}

\bmsection*{Acknowledgments}
The authors thank Avi Feller, Kenneth Frank, Erin Hartman, Melody Huang, Yaxuan Huang, Licong Lin, Sizhu Lu, and Dan Soriano for helpful comments and suggestions. The authors would also like to thank the reviewers and editorial team for their feedback which improved the quality of our manuscript. Andy Shen is partially supported by the National Science Foundation (NSF) Graduate Research Fellowship under Grant No. 2146752. Samuel D. Pimentel is supported by the NSF under Grant No. 2142146. Any opinion, findings, and conclusions or recommendations expressed in this material are those of the authors(s) and do not necessarily reflect the views of the NSF. Ran Barzilay is supported by the National Institute of Mental Health (Grant No. R21MH130797, P50MH115838) and the American Foundation for Suicide Prevention (Grant No. SRG-0-006-22). 

\bmsection*{Financial disclosure}
Ran Barzilay serves on the scientific Advisory Board for and holds equity in Taliaz Health (no conflict with the current work).

\bmsection*{Conflict of interest}
The authors declare no potential conflict of interests.

\bmsection*{Data Availability Statement}
The functions used in the analysis are available in the \texttt{decompsens} package in R: \href{https://github.com/aashen12/decompsens}{https://github.com/aashen12/decompsens}. For those with access to the ABCD data, the code used to reproduce the results can be found on GitHub: \href{https://github.com/aashen12/disparity-sensitivity}{https://github.com/aashen12/disparity-sensitivity}.

Data used in the preparation of this article were obtained from the \href{https://abcdstudy.org}{Adolescent Brain Cognitive Development (ABCD) Study}, held in the NIMH Data Archive (NDA). This is a multisite, longitudinal study designed to recruit more than 10,000 children age 9-10 and follow them over 10 years into early adulthood. The ABCD Study is supported by the National Institutes of Health and additional federal partners under award numbers U01DA041048, U01DA050989, U01DA051016, U01DA041022, U01DA051018, U01DA051037, U01DA050987, U01DA041174, U01DA041106, U01DA041117, U01DA041028, U01DA041134, U01DA050988, U01DA051039, U01DA041156,U01DA041025,U01DA041120, U01DA051038, U01DA041148, U01DA041093, U01DA041089, U24DA041123, U24DA041147. A full list of supporters is available at \href{https://abcdstudy.org/federal-partners.html}{this link}. A listing of participating sites and a complete listing of the study investigators can be found at \href{https://abcdstudy.org/consortium_members}{this link}. ABCD consortium investigators designed and implemented the study and/or provided data but did not necessarily participate in the analysis or writing of this report. This manuscript reflects the views of the authors and may not reflect the opinions or views of the NIH or ABCD consortium investigators.

\bibliography{bibliography}

\appendix

\bmsection{Proofs and Derivations} \label{sec:proofs}
\bmsubsection{Identification of counterfactual mean} \label{sec:appendix-mu10}

Under Assumptions \ref{as:condig-paper} and \ref{as:overlap-paper} (as well as SUTVA), the counterfactual outcome term $\muint$ under the stochastic intervention of $Z$ can be identified as
\begin{align*}
    \mu_{R_{0}} &\coloneqq \E\brackets{Y(R_{0}) \mid G = 1}\\
    &= \int_{X}\E\brackets{Y(R_{0}) \mid G = 1, X} f\left(X \mid G = 1\right) dX\\
    &= \int_{X}\sum_Z \E\brackets{Y(z) \mid G = 1, X} P(Z=z \mid G = 0, X) f\left(X \mid G = 1 \right) dX\\
    &=\int_{X} \sum_Z \E\brackets{Y(z) \mid G = 1, X, Z} P(Z=z \mid G = 0, X) f\left(X  \mid G = 1 \right) dX \\
    &= \int_{X} \sum_Z \E\brackets{Y(z) \mid G = 1, X, Z} P(Z=z \mid G = 1, X) \underbrace{\frac{P(Z=z \mid G = 0, X)}{P(Z=z \mid G = 1, X)}}_{\mathrm{RMPW}} f\left(X \mid G = 1 \right) dX\\
    &= \int_{X} \left\lbrace\E\brackets{Y(1) \mid G = 1, X, Z = 1} P(Z=1 \mid G = 1, X)\frac{e_0}{e_1} +  \right.\\
    &\left.\qquad \E\brackets{Y(0) \mid G = 1, X, Z=0} P(Z=0 \mid G = 1, X) \frac{1-e_0}{1-e_1} \right\rbrace f\left(X \mid G = 1 \right) dX\\
    &= \int_{X} \left\lbrace\E\brackets{Y(1) \mid G = 1, X} \E\left[Z\mid G = 1, X\right]\frac{e_0}{e_1} +  \right.\\
    &\left.\qquad \E\brackets{Y(0) \mid G = 1, X} \E\left[1-Z\mid G = 1, X\right] \frac{1-e_0}{1-e_1} \right\rbrace f\left(X \mid G = 1 \right) dX \\
    &= \int_{X} \E\brackets{\frac{e_0}{e_1} ZY(1) + \frac{1-e_0}{1-e_1} (1-Z)Y(0) \mid G = 1, X} f\left(X \mid G = 1 \right) dX\\
    &= \E\brackets{\frac{e_0}{e_1} ZY(1) + \frac{1-e_0}{1-e_1} (1-Z)Y(0) \mid G = 1}\\
    &= \E\brackets{\frac{e_0}{e_1} ZY + \frac{1-e_0}{1-e_1} (1-Z)Y \mid G = 1}\\
    &= \E\brackets{w Y \mid G = 1},
\end{align*}
where the 3rd equality follows from the fact that $R_{Z \mid G=0, X}$ is a random draw from the distribution $P\left(Z \mid G=0, X\right)$, and the 4th and 7th equalities follow from Assumption \ref{as:condig-paper}. Also recall $w = \wrmpw$.

\newpage

\bmsubsection{Identification of counterfactual mean under allowability} \label{sec:appendix-mu10-allow}

Under the allowability framework and Assumptions \ref{as:condig-paper} and \ref{as:overlap-paper} (as well as SUTVA), the counterfactual outcome term $\muint$ under the stochastic intervention of $Z$ can be identified as
\begin{align*}
    \mu_{R_{0}^{a}} &\coloneqq \E\brackets{Y(R_{0}) \mid G = 1}\\
    &= \int_{X^A} \sum_Z \E\brackets{Y(R_{0}) \mid G = 1, X^A} f\left(X^A \mid G = 1\right) dX^A\\
    &= \int_{X^A} \sum_Z \E\brackets{Y(z) \mid G = 1, X^A} P(Z=z \mid G = 0, X^A) f\left(X^A \mid G = 1 \right) dX^A\\
    &=\int_{X^A} \int_{X^N} \sum_Z \E\brackets{Y(z) \mid G = 1, X^A, X^N, Z} P(Z=z \mid G = 0, X^A)  \\
    &\qquad f\left(X^N\mid X^A, G = 1 \right) f\left(X^A\mid G = 1 \right) dX^N dX^A \\
    &= \int_{X^A} \int_{X^N} \sum_Z \E\brackets{Y(z) \mid G = 1, X^A, X^N, Z} P(Z=z \mid G = 1, X^A, X^N) \underbrace{\frac{P(Z=z \mid G = 0, X^A)}{P(Z=z \mid G = 1, X^A, X^N)}}_{\mathrm{RMPW}} \\
    &\qquad f\left(X^A, X^N \mid G = 1 \right) dX^N dX^A\\
    &= \int_{X^A} \int_{X^N} \left\lbrace\E\brackets{Y(1) \mid G = 1, X^A, X^N, Z = 1} P(Z=1 \mid G = 1, X^A, X^N)\frac{e_{0a}}{e_1} +  \right.\\
    &\left.\qquad \E\brackets{Y(0) \mid G = 1, X^A, X^N, Z=0} P(Z=0 \mid G = 1, X^A, X^N) \frac{1-e_{0a}}{1-e_1} \right\rbrace f\left(X^A, X^N \mid G = 1 \right) dX^N dX^A\\
    &= \int_{X^A} \int_{X^N} \left\lbrace\E\brackets{Y(1) \mid G = 1, X^A, X^N} \E\left[Z\mid G = 1, X^A, X^N\right]\frac{e_{0a}}{e_1} +  \right.\\
    &\left.\qquad \E\brackets{Y(0) \mid G = 1, X^A, X^N} \E\left[1-Z\mid G = 1, X^A, X^N\right] \frac{1-e_{0a}}{1-e_1} \right\rbrace f\left(X^A, X^N \mid G = 1 \right) dX^N dX^A \\
    &= \int_{X^A} \int_{X^N} \E\brackets{\frac{e_{0a}}{e_1} ZY(1) + \frac{1-e_{0a}}{1-e_1} (1-Z)Y(0) \mid G = 1, X^A, X^N} f\left(X^A, X^N \mid G = 1 \right) dX^N dX^A\\
    &= \E\brackets{\frac{e_{0a}}{e_1} ZY(1) + \frac{1-e_{0a}}{1-e_1} (1-Z)Y(0) \mid G = 1}\\
    &= \E\brackets{\frac{e_{0a}}{e_1} ZY + \frac{1-e_{0a}}{1-e_1} (1-Z)Y \mid G = 1}\\
    &= \E\brackets{w Y \mid G = 1},
\end{align*}
where the 3rd equality follows from the fact that $R_{Z \mid G=0, X^A}$ is a random draw from the distribution $P\left(Z \mid G=0, X^A\right)$, and the 4th and 7th equalities follow from Assumption \ref{as:condig-paper}. Also recall $w = \wrmpwallow$.

\newpage

\bmsubsection{Proof of Theorem \ref{thm:valid-pb}} \label{sec:proof-valid-pb}

%\label{prf:valid-pb}
\newcommand{\phigrad}{\dot{\Phi}_0}
\newcommand{\XAT}{X^{A'}}

An outline of the proof is as follows. We first express the counterfactual outcome $\muint$, our causal estimand of interest, in the $Z$-estimation framework. From there, we use the asymptotic theory of bootstrap for $Z$-estimators to derive the validity of the percentile bootstrap. Note that a similar form of the proof can be found in Zhao et al. (2019)\cite{zhao2019sensitivity} where the weights are traditional inverse propensity weights instead of RMPW weights. Like Zhao et al. (2019)\cite{zhao2019sensitivity}, we also assume the propensity scores that comprise the RMPW weight follow a logistic model. 

To begin, we define $e_0 = P\paren{Z=1 \mid X^A, G = 0}$ and $e_1 = P\paren{Z=1 \mid X, G = 1}$. Since we assume the weights follow a logistic model, we have that
\begin{align*}
    e_g &= \frac{e^{\beta_g' X}}{1+e^{\beta_g' X}}\\
    e_g^{\ast} &= \frac{e^{\beta_g^{\ast'} X}}{1+e^{\beta_g^{\ast'} X}},
\end{align*}
where $\beta_g \in \R^{p}$ represents the logistic regression coefficient for group $g \in \braces{0,1}$ and $\beta_g^{\ast}$ represents the corresponding true parameter value. Since $e_0$ is only dependent on allowable covariates, $\beta_0$ and $\betaast_0$ are defined only for the first $p_A$ coordinates (denoting allowable covariates) and the remaining $p_N$ coordinates are 0 (denoting non-allowable covariates that are not conditioned on).

\newcommand{\muwh}{\mu_w^{(h)}}
\newcommand{\muh}{\mu^{(h)}}

Next, define the following parameters:
\begin{align*}
    \muwh &= \E\brackets{G e^{h(X,U)} \paren{\frac{e_0^{\ast}}{e_1^{\ast}}Z + \frac{1-e_0^{\ast}}{1-e_1^{\ast}}(1-Z)} }\\
    \muh &= \frac{1}{\muwh} \E\brackets{GY e^{h(X,U)}\paren{\frac{e_0^{\ast}}{e_1^{\ast}}Z + \frac{1-e_0^{\ast}}{1-e_1^{\ast}}(1-Z)}}.
\end{align*}
Our parameter vector is then $\theta = \paren{\mu, \mu_w, \beta_0, \beta_1}' \in \Theta$, where $\theta \in \R^{2 + p_A + p_N}$. We define the true parameter vector as $\theta^{\ast} = \paren{\muh, \muwh, \betaast_0, \betaast_1}$.

For $t = \paren{g, z, x^A, x^N, y}' \in \braces{0,1} \times \braces{0, 1} \times \R^{p_A} \times \R^{p_N} \times \R$, define the function $Q : \braces{0,1} \times \braces{0, 1} \times \R^{p_A} \times \R^{p_N} \times \R \mapsto \R^{p_A + p_N + 2}$, where 
\begin{align*}
    \Qttheta = \begin{pmatrix}
        Q_1\paren{t \mid \theta}\\
        Q_2\paren{t \mid \theta}\\
        Q_3\paren{t \mid \theta}\\
        Q_4\paren{t \mid \theta}
    \end{pmatrix} \coloneqq
    \begin{pmatrix}
        \paren{Z - \expitwhitelower} x (1-g)\\
        \paren{Z - \expitblacklower} x g\\
        \mu_w - g~e^{h(X,U)}\brackets{e^{\betawhitelower - \betablacklower} \frac{1+\expblacklower}{1+\expwhitelower}z + \frac{1+\expblacklower}{1+\expwhitelower}\paren{1-z}}\\
        \mu_w \mu - gy~e^{h(X,U)}\brackets{e^{\betawhitelower - \betablacklower} \frac{1+\expblacklower}{1+\expwhitelower}z + \frac{1+\expblacklower}{1+\expwhitelower}\paren{1-z}}
    \end{pmatrix}.
\end{align*}
Note that each element of $\Qttheta$ is the same as those in Zhao et al. (2019)\cite{zhao2019sensitivity}, except we use a RMPW weight which requires estimating a ratio of two propensity scores instead of a single propensity score in the traditional IPW case of Zhao et al. (2019)\cite{zhao2019sensitivity}, resulting in an extra function for the additional logistic regression coefficient. As such, we prove that asymptotic normality of the bootstrapped $Z$-estimators still holds in this regime. 

Then, define
\begin{align*}
    \Phi(\theta) = \int \Qttheta d\mathbb{P}^{\ast}(t),
\end{align*}
where $T = \paren{G,Z, X, Y}' \sim \mathbb{P}^{\ast}$ and $\mathbb{P}^{\ast}$ is the true data-generating distribution of $T$. Observe that $\Phi\left(\theta^{\ast}\right) = 0$ when $\theta = \theta^{\ast}$, the true parameter values.

One can obtain the $Z$-estimates $\paren{\hat{\mu}^{(h)}, \hat{\mu}_w^{(h)}, \hat{\beta}_0, \hat{\beta}_1}$ by solving the following system of equations:
\begin{align*}
    \hat{\Phi}_n \paren{\theta} &\coloneqq \ninvsumin Q\paren{t_i \mid \theta} =\\
    &\begin{pmatrix}
        \ninvsumin\brackets{Z_i - \frac{\expwhitehat}{1+\expwhitehat}} X_i \paren{1-G_i}\\
        \ninvsumin\brackets{Z_i - \frac{\expblackhat}{1+\expblackhat}} X_i G_i\\
        \hat{\mu}_w^{(h)} - \ninvsumin G_i e^{h(X_i, Y_i)} \brackets{e^{\betawhitehat - \betablackhat} \frac{1+\expblackhat}{1+\expwhitehat}Z_i + \frac{1+\expblackhat}{1+\expwhitehat}\paren{1-Z_i}}\\
        \hat{\mu}_w^{(h)} \hat{\mu}^{(h)} - \ninvsumin G_i Y_i e^{h(X_i, Y_i)} \brackets{e^{\betawhitehat - \betablackhat} \frac{1+\expblackhat}{1+\expwhitehat}Z_i + \frac{1+\expblackhat}{1+\expwhitehat}\paren{1-Z_i}}
    \end{pmatrix} =0 
\end{align*}    
It is easy to see that $\hat{\mu}^{(h)}$ is precisely the RMPW estimate of $\muint$ and $\hat{\beta}_g$ for $g \in \braces{0,1}$ are the respective maximum likelihood estimates for the logistic regression propensity score models under the allowability framework.

Finally, define $\dot{\Phi}_0 = \E\brackets{\nabla_{\theta = \theta^{\ast}} \Qttheta}$ and $\Sigma = \E\brackets{\Qttheta \Qttheta'}$. Zhao et al. (2019)\cite{zhao2019sensitivity} and Huang and Pimentel (2024)\cite{huang2024variance} invoke a set of regularity assumptions in order to verify asymptotic normality of their bootstrapped $Z$-estimators. We invoke an equivalent set of assumptions with an additional assumption for estimating two propensity scores in the weights ($e_0$ and $e_1$) as opposed to a single propensity score in Zhao et al. (2019)\cite{zhao2019sensitivity} and Huang and Pimentel (2024)\cite{huang2024variance}:

\begin{assumption}[Regularity Assumptions]
\label{as:regularity-pb}
    Assume the parameter space $\Theta$ is compact and the true parameter $\theta^{\ast}$ is in the interior of $\Theta$. 
    %\red{\textbf{Assume further that $X$ has compact support.}} 
    Moreover, the joint distribution of $(Y, X^A, X^N) = (Y, X)$ satisfies
    \begin{enumerate}
        \item $\E\paren{Y^4} < \infty$.
        \item $\left|\det\paren{\E\brackets{\frac{e^{\beta_g^{\ast '}}}{\paren{1+e^{\beta_g^{\ast '}}}^2} X X'}}\right| > 0$ for all $g \in \braces{0,1}$.
        %\item $\left|\det\paren{\E\brackets{\frac{\expblackast}{\paren{1+\expblackast}^2} X X'}}\right| > 0$.
        %\item The covariate vector $X$ has compact support.
        %\item \andy{I will add the supremum assumption once I figure out what the analog of the assumption even is in our case.}
        \item $\E\brackets{\sup_{\beta_g \in S} e^{\beta_g'X}} < \infty$
        for every compact subset $S \in \R^{p}$ and for all $g \in \braces{0,1}$.
    \end{enumerate}
\end{assumption}

%Under this assumption, we follow \cite{kosorok2008introduction} and Zhao et al. (2019)\cite{zhao2019sensitivity} and state the limiting distribution of $Z$-estimators:

First, we show that $\phigrad$ and $\Sigma$ are well-defined. A direct computation gives 
\begin{align*}
      \phigrad = \scriptsize \begin{pmatrix}
        0 & 0 & -\E{\frac{\expwhiteast}{\paren{1+\expwhiteast}^2} X X'} & 0\\
        0 & 0 & 0 & -\E{\frac{\expblackast}{\paren{1+\expblackast}^2} X X'}\\
        0 & 1 & \E \frac{G e^{h(X,U)}\expwhiteast\paren{1+\expblackast}}{\paren{1+\expwhiteast}^2} X' \brackets{Ze^{-\betablackast} - (1-Z)} & \E \frac{G e^{h(X,U)}}{1+\expwhiteast} X' \brackets{\paren{1-Z} \expwhiteast - e^{\betawhiteast - \betablackast} Z}\\
        \muwh & \muh & \E \frac{GY e^{h(X,U)}\expwhiteast\paren{1+\expblackast}}{\paren{1+\expwhiteast}^2} X' \brackets{Ze^{-\betablackast} - (1-Z)} & \E \frac{GY e^{h(X,U)}}{1+\expwhiteast} X' \brackets{\paren{1-Z} \expblackast - e^{\betawhiteast - \betablackast} Z}
    \end{pmatrix},
\end{align*}
and 
\begin{align*}
    \left| \det\paren{\phigrad} \right| &= \left|\det\paren{\E{\frac{\expwhiteast}{\paren{1+\expwhiteast}^2} X X'}} \det
     \begin{pmatrix}
        0 & 0 & -\E{\frac{\expblackast}{\paren{1+\expblackast}^2} X X'}\\
        0 & 1 & \E \frac{G e^{h(X,U)}}{1+\expwhiteast} X' \brackets{\paren{1-Z} \expblackast - e^{\betawhiteast - \betablackast} Z}\\
        \mu_w & \mu & \E \frac{GY e^{h(X,U)}}{1+\expwhiteast} X' \brackets{\paren{1-Z} \expblackast - e^{\betawhiteast - \betablackast} Z}
    \end{pmatrix}
    \right|\\
    &= \left|\mu_w \det\paren{\E{\frac{\expwhiteast}{\paren{1+\expwhiteast}^2} X X'}} \det\paren{\E{\frac{\expblackast}{\paren{1+\expblackast}^2} X X'}}\right| > 0,
\end{align*}
by Assumption \ref{as:regularity-pb} (2). Moreover, we have that $\Sigma < \infty$ by Assumption \ref{as:regularity-pb} (1) and direct multiplication. Therefore $\phigrad$ is invertible and well-defined.

Proving the asymptotic normality of bootstrapped $Z$-estimators requires verifying the following conditions \citep{kosorok2008introduction,zhao2019sensitivity,huang2024variance}:
\begin{enumerate}
\item The class of functions \( \braces{t \mapsto Q\left(t\mid\theta\right) : \theta \in \Theta} \) is \( \mathbb{P} \)-Glivenko-Cantelli.
\item \( \left\|\Phi\left(\theta\right)\right\|_1 \) is strictly positive outside every open neighborhood of \( \theta^{\ast} \).
\item The class of functions \( \braces{t \mapsto Q\left(t\mid\theta\right) : \theta \in \Theta} \) is \( \mathbb{P} \)-Donsker, and \( \mathbb{E}\left[\left(Q\left(T|\theta_n\right) - Q\left(T|\theta^{\ast}\right)\right)^2\right] \to 0 \) whenever \( \left\|\theta_n - \theta^{\ast}\right\|_1 \to 0 \).
\end{enumerate}

\paragraph{Condition 1:}The class of functions \( \braces{t \mapsto Q\left(t\mid\theta\right) : \theta \in \Theta} \) is \( \mathbb{P} \)-Glivenko-Cantelli.

\begin{proof}
    Define the envelope function $B\paren{t \mid \theta} = \sup_{\theta \in \Theta} \bignorm{\Qttheta}_1$. Notice that $\infnorm{h(X,U)} \leq \gamma$ and $\bigabs{g} \leq 1$. Using the compactness of $\Theta$, we have the following:
\begin{align*}
    \bignorm{\Qttheta}_1 &\leq \bignorm{\Qittheta{1}}_1 + \bignorm{\Qittheta{2}}_1 + \bigabs{\Qittheta{3}} + \bigabs{\Qittheta{4}}\\
    &\leq 2 \bignorm{x}_1 + \bigabs{\mu_w} + \bigabs{\mu_w \mu} + e^{\gamma}\brackets{e^{\betawhitelower - \betablacklower} \frac{1+\expblacklower}{1+\expwhitelower}z + \frac{1+\expblacklower}{1+\expwhitelower}\paren{1-z}}\paren{1 + \bigabs{y}} \\
    &\leq 2 \bignorm{x}_1 + e^{\gamma}\brackets{e^{\betawhitelower - \betablacklower} \frac{1+\expblacklower}{1+\expwhitelower}z + \frac{1+\expblacklower}{1+\expwhitelower}\paren{1-z}}\paren{1 + \bigabs{y}} + M,
\end{align*}
    for some absolute constant $M$. Applying Assumption \ref{as:regularity-pb} and the result above, it follows that $\E\brackets{B(T)} < \infty$ and by Wellner (2005)\cite{wellner2005empirical} Lemma 6.1, the desired class of functions is \( \mathbb{P} \)-Glivenko-Cantelli.
\end{proof}

\paragraph{Condition 2:}\(\left\|\Phi\left(\theta\right)\right\|_1 \) is strictly positive outside every open neighborhood of \( \theta^{\ast} \).

\begin{proof}
     First assume $\bignorm{\betaast_g - \beta_g}_1 > \epsilon / M$ for $g = \braces{0, 1}$. It can be shown in Zhao et al. (2019)\cite{zhao2019sensitivity} that $\left\|\Phi\left(\theta\right)\right\|_1 > 0$ by leveraging Assumption \ref{as:regularity-pb} (2).

Next, assume $\bignorm{\betaast_g - \beta_g}_1 \leq \epsilon / M$ for $g \in \braces{0, 1}$. This implies $\infnorm{\betaast_g - \beta_g} \leq \epsilon / M$. We observe that the gradient of $w$ with respect to $\beta_0'x$ and $\beta_1'x$ can be expressed as
\begin{align*}
    \nabla w &= \begin{pmatrix}
        \frac{G e^{h(X,U)}\expwhite\paren{1+\expblack}}{\paren{1+\expwhite}^2} \brackets{Ze^{-\betablack} - (1-Z)} \\ \frac{G e^{h(X,U)}}{1+\expwhite} \brackets{\paren{1-Z} \expblack - e^{\betawhite - \betablack} Z}
    \end{pmatrix} < \infty,
\end{align*}
where the inequality follows from the compactness of $\Theta$ and Assumption \ref{as:regularity-pb} (2). As a result of the compactness of $\Theta$ and Assumption \ref{as:regularity-pb}, the gradient of $w_i$ is well-defined and we obtain an upper bound on $\bigabs{w - \wast}$ using the mean value theorem:
    \begin{align*}
       \bigabs{w - \wast} &= \bigabs{\nabla w_{\paren{\beta_0'x, \beta_1'x} = \left(c_0, c_1 \right)}' \begin{pmatrix}
            \paren{\beta_{0} - \betaast_{0}}' x\\
            \paren{\beta_{1} - \betaast_{1}}' x
        \end{pmatrix}} \quad\text{(for some $c_j \in \brackets{\beta_{j}'x, \beta_{j}^{\ast '}x}, j \in \braces{0,1}$)}\\
        &\leq {\infnorm{\nabla w_{\paren{\beta_0'x, \beta_1'x} = \left(c_0, c_1 \right)}}}
        \bignorm{\begin{pmatrix}
            \paren{\beta_{0} - \betaast_{0}}' x\\
            \paren{\beta_{1} - \betaast_{1}}' x
        \end{pmatrix}}_1 \\
        &\lesssim {\bigabs{\paren{\beta_{0} - \betaast_{0}}' x} + \bigabs{\paren{\beta_{1} - \betaast_{1}}' x}}\\
        &\leq {\bignorm{x}_2\paren{\bignorm{\beta_{0} - \betaast_{0}}_2 + \bignorm{\beta_{1} - \betaast_{1}}_2}}\\
        &\leq {\bignorm{x}_1\paren{\bignorm{\beta_{0} - \betaast_{0}}_1 + \bignorm{\beta_{1} - \betaast_{1}}_1}},
    \end{align*}
    where the first inequality follows from H{\"o}lder's inequality and the third from Cauchy-Schwarz.

From there, we have the following:
\begin{align*}
    \bigabs{\E\brackets{Ge^{h(X,U)+ \log(\wast)}  - Ge^{h(X,U) + \log(w)}}} &\lesssim \E\brackets{\bigabs{Ge^{h(X,U)}} \cdot \bigabs{\wast - w}}\\
    %&\leq e^{\gamma} \infnorm{\betaast - \beta} \E\brackets{\bignorm{X}_1}\\
    &\lesssim \paren{ \infnorm{\betaast_0 - \beta_0} + \infnorm{\betaast_1 - \beta_1}} e^{\gamma} \E\brackets{\bignorm{X}_1}\\
    & \leq K_1\paren{\gamma} \frac{2\varepsilon}{M} \leq \frac{\varepsilon}{64K},
\end{align*}
by following Zhao et al. (2019)\cite{zhao2019sensitivity} and choosing 
\begin{align*}
    M \geq 128K \cdot K_1\paren{\gamma},
\end{align*}
where
\begin{align*}
    K_1\paren{\gamma} &\coloneqq e^{\gamma} \E\brackets{\bignorm{X}_1} < \infty\text{, and}\\
    K &\coloneqq \sup_{\mu \in \Theta} \bigabs{\mu} \in \paren{0, \infty}
\end{align*}
by Assumption \ref{as:regularity-pb} and the compactness of $\Theta$.

It then follows that whenever $\bignorm{\beta_g - \betaast_g}_{1} \leq \varepsilon/K$ for $g = 0, 1$ and $\bigabs{\mu_w - \muwh} > \frac{\varepsilon}{4K}$,
\begin{align*}
    \left\|\Phi\left(\theta\right)\right\|_1 \geq \bigabs{\mu_w - \muwh  + \E\brackets{Ge^{h(X,U)+ \log(\wast)}  - Ge^{h(X,U) + \log(w)}}} > 0.
\end{align*}

Finally, assume $\bignorm{\beta_g - \betaast_g}_{\infty} \leq \varepsilon/K$ for $g = 0, 1$. It follows that
\begin{align*}
    \bigabs{\E\brackets{GYe^{h(X,U)+ \log(\wast)}  - GYe^{h(X,U) + \log(w)}}} &\lesssim \E\brackets{\bigabs{Ge^{h(X,U)}} \cdot \bigabs{e^{\log(\wast)} - e^{\log(w)}}}\\
    &=\E\brackets{\bigabs{Ge^{h(X,U)}} \cdot \bigabs{YX'\paren{\paren{\beta_0 - \betaast_0} + \paren{\beta_1 - \betaast_1}}}}\\
    %&\leq e^{\gamma} \infnorm{\betaast - \beta} \E\brackets{\bignorm{X}_1}\\
    &\leq \paren{ \infnorm{\betaast_0 - \beta_0} + \infnorm{\betaast_1 - \beta_1}} e^{\gamma} \E\brackets{\bignorm{YX}_1}\\
    & \leq K_2\paren{\gamma} \frac{2\varepsilon}{M} \leq \frac{\varepsilon}{64K},
\end{align*}
by choosing $ M \geq 128K \cdot K_2\paren{\gamma}$
where
\begin{align*}
    K_2\paren{\gamma} &\coloneqq e^{\gamma} \E\brackets{\bignorm{YX}_1} < \infty,
\end{align*}
by Assumption \ref{as:regularity-pb}. Then, for $\bignorm{\mu_w - \muwh}_{1} \leq \frac{\varepsilon}{4K}$, it follows from the definition of $K$ that $\bignorm{\mu_w \mu - \muwh \mu}_1 \leq \frac{\varepsilon}{4}$. Then, for $\bignorm{\mu -\muh}_1 > \frac{\varepsilon}{2\muwh}$, we have the following:
\begin{align*}
    \left\|\Phi\left(\theta\right)\right\|_1 &\geq \bigabs{\mu_w \mu - \muwh \muh + \E\brackets{GYe^{h(X,U)+ \log(\wast)}  - GYe^{h(X,U) + \log(w)}}} \\
    \quad &= \bigabs{\mu_w\mu - \muwh\mu +\muwh\mu - \muwh\muh +\E\brackets{GYe^{h(X,U)+ \log(\wast)}  - GYe^{h(X,U) + \log(w)}}}\\
    \quad &= \bigabs{\mu_w\mu - \muwh\mu +\muwh\paren{\mu -\muh}+\E\brackets{GYe^{h(X,U)+ \log(\wast)}  - GYe^{h(X,U) + \log(w)}}} > 0.
\end{align*}
Combining these three equations, it follows that $\left\|\Phi\left(\theta\right)\right\|_1 > 0$ for every open neighborhood of $\theta^{\ast}$. 
\end{proof}

\paragraph{Condition 3:}The class of functions \( \braces{t \mapsto Q\left(t\mid\theta\right) : \theta \in \Theta} \) is \( \mathbb{P} \)-Donsker, and \( \mathbb{E}\left[\left(Q\left(T|\theta_n\right) - Q\left(T|\theta^{\ast}\right)\right)^2\right] \to 0 \) whenever \( \left\|\theta_n - \theta^{\ast}\right\|_1 \to 0 \).

\begin{proof}

    Let 
    \begin{align*}
        \theta_1 &= \paren{\mu_1, \mu_{w1}, \beta_{01}, \beta_{11}}\\
        \theta_2 &= \paren{\mu_2, \mu_{w2}, \beta_{02}, \beta_{12}}
    \end{align*}
    be two points in the parameter space $\Theta$. Using the mean value Theorem and results from Zhao et al. (2019)\cite{zhao2019sensitivity}, we have that
    \begin{align*}
        \bignorm{Q_1\paren{t \mid \theta_2} - Q_1\paren{t \mid \theta_1}}_1 &\lesssim M_1(x) \bignorm{\beta_{02} - \beta_{01}}_1,\text{ and}\\
        \bignorm{Q_2\paren{t \mid \theta_2} - Q_2\paren{t \mid \theta_1}}_1 &\lesssim M_1(x) \bignorm{\beta_{12} - \beta_{11}}_1,
    \end{align*}
    where $M_1(x) = \bignorm{x}_1^2$.

    Next, recall that the weight $w_i$ is defined as a function of $\beta_{0i}'x$ and $\beta_{1i}'x$. As a result of the compactness of $\Theta$ and Assumption \ref{as:regularity-pb}, the gradient of $w_i$ is well-defined and we obtain an upper bound on $\bigabs{w_2 - w_1}$ using the mean value Theorem:
    \begin{align*}
        \bigabs{w_2 - w_1} &= \bigabs{\nabla w_{\paren{\beta_0'x, \beta_1'x} = (c_0, c_1)}' \begin{pmatrix}
            \paren{\beta_{02} - \beta_{01}}' x\\
            \paren{\beta_{12} - \beta_{11}}' x
        \end{pmatrix}} \quad\text{(for some $c_j \in \brackets{\beta_{j1}'x, \beta_{j2}'x}, j \in \braces{0,1}$)}\\
        &\leq {\infnorm{\nabla w_{\paren{\beta_0'x, \beta_1'x} = (c_0, c_1)}}}
        \bignorm{\begin{pmatrix}
            \paren{\beta_{02} - \beta_{01}}' x\\
            \paren{\beta_{12} - \beta_{11}}' x
        \end{pmatrix}}_1 \quad\text{(H{\"o}lder's inequality)}\\
        &\lesssim {\bigabs{\paren{\beta_{02} - \beta_{01}}' x} + \bigabs{\paren{\beta_{12} - \beta_{11}}' x}}\\
        &\leq {\bignorm{x}_2\paren{\bignorm{\beta_{02} - \beta_{01}}_2 + \bignorm{\beta_{12} - \beta_{11}}_2}}\quad\text{(Cauchy-Schwarz inequality)}\\
        &\leq {\bignorm{x}_1\paren{\bignorm{\beta_{02} - \beta_{01}}_1 + \bignorm{\beta_{12} - \beta_{11}}_1}},
    \end{align*}
    where the first inequality follows from H{\"o}lder's inequality and the third from Cauchy-Schwarz. Thus, we have the following:
    \begin{align*}
        \bignorm{Q_3\paren{t \mid \theta_2} - Q_3\paren{t \mid \theta_1}}_1 &\lesssim \bigabs{\mu_{w2} - \mu_{w1}} + e^{\gamma} \bigabs{w_2 - w_1}\\
        &\lesssim \bigabs{\mu_{w2} - \mu_{w1}} + {\bignorm{x}_1\paren{\bignorm{\beta_{02} - \beta_{01}}_1 + \bignorm{\beta_{12} - \beta_{11}}_1}}\\
        &\lesssim M_2(x) \paren{\bigabs{\mu_{w2} - \mu_{w1}} + \bignorm{\beta_{02} - \beta_{01}}_1 + \bignorm{\beta_{12} - \beta_{11}}_1}
    \end{align*}
    where $M_2(x)= 1+ \bignorm{x}_1$. 

    Finally, using the bound on $\bigabs{w_2 - w_1}$, we obtain the following:
    \begin{align*}
        \bignorm{Q_4\paren{t \mid \theta_2} - Q_4\paren{t \mid \theta_1}}_1 &\lesssim \bigabs{\mu_{w2}\mu_2 - \mu_{w1}\mu_1} + e^{\gamma} \bigabs{y\paren{w_2 - w_1}}\\
        &\lesssim \bigabs{\mu_{w2} - \mu_{w1}} + \bigabs{\mu_2 - \mu_1} + \bignorm{yx}_1 \paren{\bignorm{\beta_{02} - \beta_{01}}_1 + \bignorm{\beta_{12} - \beta_{11}}_1}\\
        &\lesssim M_3(x) \paren{\bigabs{\mu_{w2} - \mu_{w1}} + \bigabs{\mu_2 - \mu_1} + \bignorm{\beta_{02} - \beta_{01}}_1 + \bignorm{\beta_{12} - \beta_{11}}_1},
    \end{align*}
    where $M_3(x,y)= 1+ \bignorm{yx}_1$.

    Defining $M(x,y) = 2M_1(x) + M_2(x) + M_3(x,y)$ and combining the previous three results yields
    \begin{align*}
        \bignorm{Q\paren{t \mid \theta_2} - Q\paren{t \mid \theta_1}}_1 &= \sum_{k=1}^4 \bignorm{Q_k\paren{t \mid \theta_2} - Q_k\paren{t \mid \theta_1}}_1\\
        &\lesssim M(x,y) \bignorm{\theta_2 - \theta_1}_1.
    \end{align*}
    Moreover, since $\E\brackets{M(x,y)^2} < \infty$ by Assumption \ref{as:regularity-pb}, this we have shown that the set of functions \( \braces{t \mapsto Q\left(t\mid\theta\right) : \theta \in \Theta} \) is Lipschitz, and thus \( \mathbb{P} \)-Donsker. Therefore, it must also hold that
    \( \mathbb{E}\left[\left(Q\left(T|\theta_n\right) - Q\left(T|\theta^{\ast}\right)\right)^2\right] \to 0 \) whenever \( \left\|\theta_n - \theta^{\ast}\right\|_1 \to 0 \).
\end{proof}
With these three conditions verified, one can then apply Kosorok (2008)\cite{kosorok2008introduction}, Theorem 10.16:
\begin{align*}
    \sqrt{n}(\hat{\theta} - \theta) \overset{d}{\to} N \left( 0, \phigrad^{-1}\Sigma \phigrad\right), \quad \text{and} \quad \sqrt{n}(\hat{\hat{\theta}} - \theta) \overset{d}{\to} N \left( 0, \phigrad^{-1}\Sigma \phigrad \right).
\end{align*}
Applying the Delta method gives the following corollary:
\begin{align*}
    \sqrt{n} \left( \hat{\mu}^{(h)} - \mu^{(h)} \right) \overset{d}{\to} N\left(0, \left(\sigma^{(h)}\right)^2\right), \quad \text{and} \quad \sqrt{n} \left( \hat{\hat{\mu}}^{(h)} - \hat{\mu}^{(h)} \right) \overset{d}{\to} N\left(0, \left(\sigma^{(h)}\right)^2\right),
\end{align*}
where $\left(\sigma^{(h)}\right)^2 = \left(\phigrad^{-1}\Sigma \phigrad\right)_{11}$ is the first diagonal element of $\phigrad^{-1}\Sigma \phigrad$. Applying results from Zhao et al. (2019)\cite{zhao2019sensitivity} Appendix C and Huang and Pimentel (2024)\cite{huang2024variance} Appendix B.2 completes the proof. $\qed$

\newpage

\subsection{Proof of Theorem \ref{thm:amp-bias-decomp} and Corollary \ref{cor:part-id-bias}} \label{sec:proof-ignor-bias}

Since $U$ is non-allowable, the definition of $\E\brackets{Y(int) \mid G = 1, X, U}$ remains unchanged since $e_0^{\ast} = e_0$:
\begin{align*}
    \E\brackets{Y(int) \mid G = 1, X, U} &= e_0 \E\brackets{Y(1) \mid G = 1, X, U} + (1-e_0) \E\brackets{Y(0) \mid G = 1, X, U},
\end{align*}
where $\E\brackets{Y(z) \mid G = 1, X, U}$ are defined in Equations \eqref{eqn:model-y1} and \eqref{eqn:model-y0} for $z = 0, 1$.

We can express the $X, U$-conditional expectation of the RMPW estimand for group $G=1$ as
\begin{align*}
    \E\brackets{wY \mid G = 1, X, U} &= \E\brackets{w\paren{ZY(1) + (1-Z) Y(0)} \mid G = 1, X, U}\nonumber\\
    &= \E\brackets{\frac{e_0}{e_1}Z Y(1) \mid G = 1, X, U} + \E\brackets{\frac{1-e_0}{1-e_1} (1-Z) Y(0) \mid G = 1, X, U}\\
    &= \frac{e_0}{e_1} \E\brackets{Z  \mid G = 1, X, U} \E\brackets{Y(1) \mid G = 1, X, U} \\
    &\quad+ \frac{1-e_0}{1-e_1} \E\brackets{1-Z  \mid G = 1, X, U} \E\brackets{Y(0) \mid G = 1, X, U}\\
    &= \frac{e_0}{e_1} \E\brackets{Z  \mid G = 1, X, U} \paren{f(X) + \beta_z + \beta_u U} \\
    &\quad+ \frac{1-e_0}{1-e_1} \E\brackets{1-Z  \mid G = 1, X, U} \paren{f(X) + \beta_u U}\\
    &= \E\brackets{\frac{e_0}{e_1} Z f(X) \mid G = 1, X, U} + \beta_z\E\brackets{\frac{e_0}{e_1} Z \mid G = 1, X, U} + \beta_u\E\brackets{\frac{e_0}{e_1} ZU \mid G = 1, X, U}\\
    & \quad+ \E\brackets{\frac{1-e_0}{1-e_1} (1-Z) f(X) \mid G = 1, X, U} + \beta_u\E\brackets{\frac{1-e_0}{1-e_1} (1-Z) U \mid G = 1, X, U},
\end{align*}
where the third equality holds due to ignorability conditional on $X, U$.

The expectation of the RMPW estimand conditional on $G=1$ ($\E\brackets{wY \mid G = 1}$) is
\begin{align*}
    \E\brackets{\frac{e_0}{e_1} Z f(X) \mid G = 1} + \beta_z \E\brackets{\frac{e_0}{e_1} Z\mid G = 1} + \beta_u\E\brackets{\frac{e_0}{e_1}  ZU\mid G = 1}+ \\
    \quad\E\brackets{\frac{1-e_0}{1-e_1} (1-Z) f(X)\mid G = 1} + \beta_u\E\brackets{\frac{1-e_0}{1-e_1} (1-Z) U\mid G = 1}.
\end{align*}
Similarly, the expectation of $Y(int)$ conditional on $G=1$ ($\E\brackets{Y(int) \mid G = 1}$) is
\begin{align*}
    \E\brackets{e_0 f(X)\mid G = 1} + \beta_z\E\brackets{e_0\mid G = 1} + \beta_u\E\brackets{e_0  U\mid G = 1}\\
    + \E\brackets{(1-e_0) f(X)\mid G = 1} + \beta_u\E\brackets{(1-e_0) U\mid G = 1}.
\end{align*}
Using the fact that $\E\brackets{Z \mid G=1, X} = e_1$ observe the following:
\begin{itemize}
    \item $\E\brackets{\frac{e_0}{e_1} Z f(X) \mid G = 1} = \E\brackets{e_0 f(X) \mid G = 1}$
    \item $\E\brackets{\frac{1-e_0}{1-e_1} (1-Z) f(X) \mid G = 1} = \E\brackets{(1-e_0) f(X) \mid G = 1}$
    \item $\E\brackets{\frac{e_0}{e_1} Z \mid G = 1} = \E\brackets{e_0 \mid G = 1}$.
\end{itemize}
Using these facts, the unconditional bias (aka ``ignorability bias'') for group $G=1$ is
\begin{align*}
    &\E\brackets{Y(int) \mid G = 1} - \E\brackets{wY \mid G = 1}\\
    &= \beta_u\paren{\E\brackets{e_0U\mid G = 1} - \E\brackets{\frac{e_0}{e_1} ZU\mid G = 1} + \E\brackets{(1-e_0)U\mid G = 1} - \E\brackets{\frac{1-e_0}{1-e_1} (1-Z)U \mid G = 1}}\\
    &= \beta_u \paren{\E\brackets{U \mid G = 1} - \E\brackets{\paren{\frac{e_0}{e_1} Z + \frac{1-e_0}{1-e_1} (1-Z)}U \mid G = 1}}\\
    &= \beta_u \paren{\E\brackets{U \mid G = 1}  - \E\brackets{wU \mid G = 1}}\\
    &= \beta_u \underbrace{\paren{\E\brackets{\frac{e_0 - e_1}{1 - e_1} \paren{U - \frac{ZU}{e_1}}\mid G = 1}}}_{\delta_u},
\end{align*}
where the last equality follows because $1-w = \frac{e_0 - e_1}{1-e_1}\paren{1 - \frac{Z}{e_1}}$. This completes the proof. $\qed$

To see why Theorem \ref{thm:amp-bias-decomp} would not hold for an unmeasured \textit{allowable} covariate, observe that the counterfactual outcome $Y(int)$ would be defined by $e_0^{\ast}$ instead of $e_0$. When taking the bias with respect to $\beta_z$, we have
\begin{align*}
    \E\brackets{Y(int) \mid G = 1} - \E\brackets{wY \mid G = 1} &= \beta_u \delta_u + \beta_z ~\E\brackets{e_0^{\ast} - e_0\mid G = 1}.
\end{align*}
This bias is equivalent to the bias term above with an extra parameter. If $U$ is non-allowable, then $e_0^{\ast} = e_0$ and this parameter is 0. 

However, Park et al. (2023)\cite{park2023sensitivity} demonstrate that their sensitivity analysis for regression holds for unmeasured allowable confounders under the added assumption of $G \indep U \mid X^A$. In our setting, we can also use this assumption to reduce the amplification back to two parameters: 
\begin{align*}
    \E\brackets{e_0^{\ast} \mid G = 1} &= \E\brackets{e_0~\frac{P(U \mid G = 0, Z=1, X^A)}{P(U \mid G=0, X^A)}\mid G = 1}\\
    &= \E\brackets{e_0~ \E\brackets{\frac{P(U \mid G = 0, Z=1, X^A)}{P(U \mid G=0, X^A)}\mid X^A, G = 1}\mid G = 1}\\
    &= \E\brackets{e_0 \int_u \frac{P(U=u\mid G = 0, Z=1, X^A)}{P(U=u\mid G=0, X^A)} dP(U=u\mid X^A, G = 1) \mid G = 1} \\
    &= \E\brackets{e_0 \int_u \frac{P(U=u\mid G = 0, Z=1, X^A)}{P(U=u\mid X^A)} dP(U=u\mid X^A) \mid G = 1}\\
    &= \E[e_0 \mid G=1],
\end{align*}
where the first equality follows from Bayes' rule and the fourth equality uses $G \indep U \mid X^A$. %We may assume without loss of generality that $U$ is discrete. 
Therefore, $\E\brackets{e_0^{\ast} - e_0\mid G = 1} = 0$ and the ignorability bias is $\beta_u \delta_u$. If one is not willing to assume $G \indep U \mid X^A$, it is possible to use the three-parameter amplification to calibrate unobserved allowable covariates, though visualization is more difficult because two-dimensional bias curves cannot be drawn.

To prove Corollary \ref{cor:part-id-bias}, one may obtain the lower and upper bounds by subtracting the RMPW estimand from the point estimate extrema computed via the fractional linear program. 

\newpage

\bmsection{Robustness check for temporal ordering} \label{sec:app-robustness}
As mentioned in Section \ref{sec:application}, we conduct a robustness check to assess the possibility of ``reverse causation'' where one's suicidal ideation at baseline would affect their parental support at a later time point. We exclude 443 children who already responded having suicidal ideation at baseline and repeat the sensitivity analysis, computing $\Lambdaast_{0.05}$ and $\Lambdaast$. The results are shown in Table \ref{tbl:abcd-robustness}.

\begin{table}[ht] 
\centering 
\begin{tabular}{lcccc} \toprule 
 \textbf{Parameter} & \textbf{Estimate (SD)} & \textbf{95\% CI} & $\Lambda^{\ast}_{0.05}$ & $\Lambda^{\ast}$ \\ \midrule 
 Observed Disparity & 0.234 (0.019) & [0.196, 0.272] & -- & --\\ 
 Disparity Reduction & 0.04 (0.017) & [0.01, 0.08] & 1.04 & 1.11\\ 
 Residual Disparity & 0.19 (0.02) & [0.142, 0.232] & 1.60 & 1.81\\ \bottomrule 
 \end{tabular} 
 \caption{Results from robustness check excluding children with suicidal ideation at baseline. Observed disparity, disparity reduction, residual disparity, and critical sensitivity parameters and 95\% confidence intervals are presented for the ABCD Study. $\Lambda^{\ast}_{0.05}$ corresponds to the critical parameter where the \textit{confidence interval} crosses 0, and $\Lambda^{\ast}$ corresponds to the critical parameter where the \textit{point estimate} crosses 0. There are no critical sensitivity parameters for the observed disparity since it is not a causal estimand.} 
 \label{tbl:abcd-robustness} 
\end{table}

\begin{figure}[ht]
    \centering    
    \includegraphics[width=0.75\linewidth]{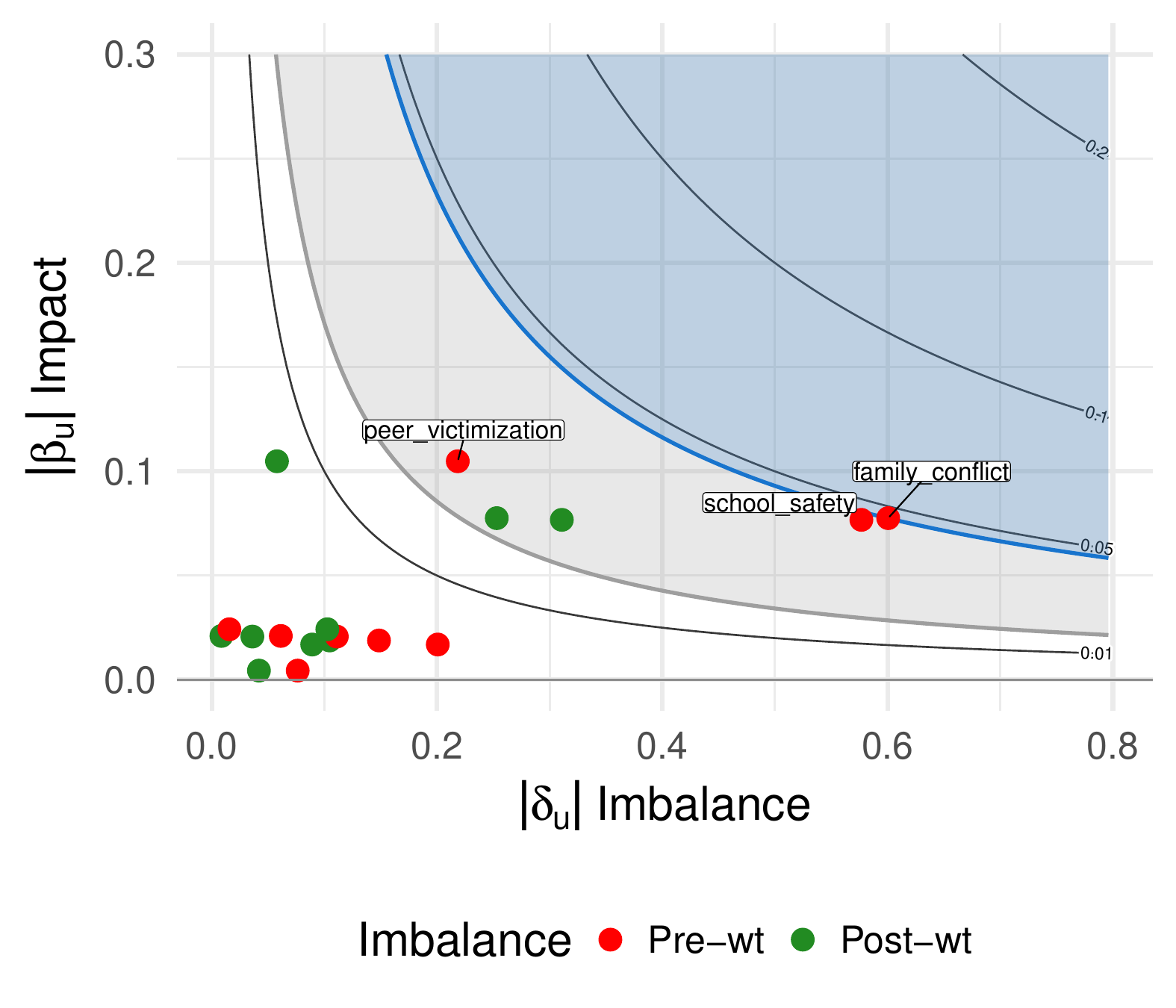}
    \caption{Bias contour plot of {\textbf{disparity reduction}} for ABCD temporal robustness check. Please refer to Figures \ref{fig:reduction-abcd-contour} and \ref{fig:residual-abcd-contour} for plot details.}
    \label{fig:reduction-abcd-robust} 
\end{figure}

\begin{figure}[ht]
    \centering
    \includegraphics[width=0.75\linewidth]{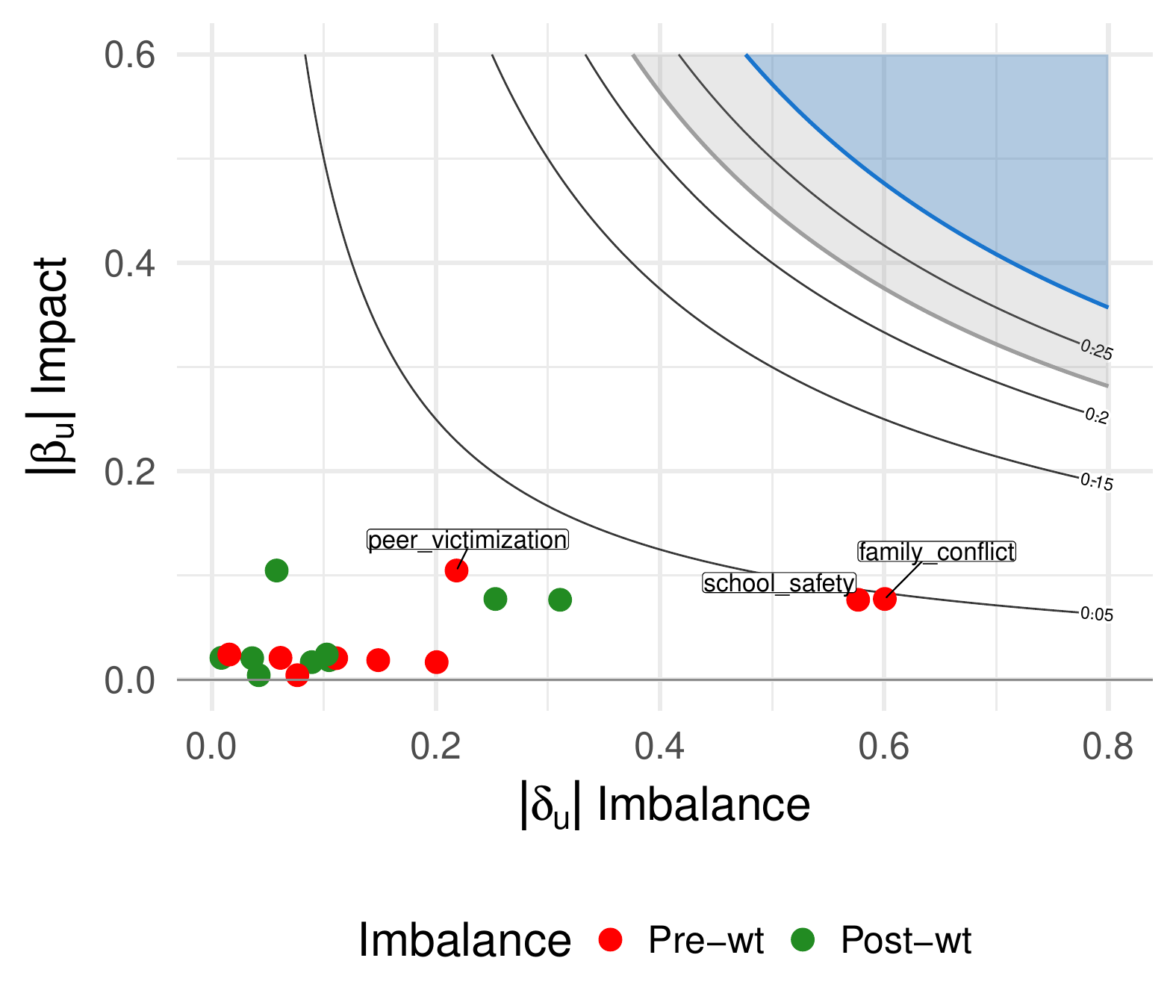}
    \caption{Bias contour plot of {\textbf{residual disparity}} for ABCD temporal robustness check. Please refer to Figures \ref{fig:reduction-abcd-contour} and \ref{fig:residual-abcd-contour} for plot details.}
    \label{fig:residual-abcd-robust}
\end{figure}

When we exclude children with suicidal ideation at baseline, we do not see a very large change in the observed disparity, the only main difference being the raw groupwise suicidal ideation rates ($\mu_1 = 0.356$, $\mu_0 = 0.122$). These values are slightly lower than those from the main analysis ($\mu_1 = 0.451$, $\mu_0 = 0.20$), but the observed disparity is very similar (0.234 vs 0.251). Moreover, the point estimate of the disparity reduction is numerically equivalent to that in Table \ref{tbl:abcd-initial-aoas} and its confidence interval is further from 0, indicating that reverse causation is highly unlikely.  In fact, the results in Table \ref{tbl:abcd-robustness} indicate that the robustness check appears to be slightly more robust to unmeasured confounding. The calibration plots for the disparity reduction and residual disparity are shown in Figures \ref{fig:reduction-abcd-robust} and \ref{fig:residual-abcd-robust}, respectively. These findings also corroborate the substantive conclusions from Section \ref{sec:application} regarding the effect of parental support on suicidal ideation.


\newpage

\begin{revisiontwo}
    
\bmsection{Robustness check for alternative definition of $Z$} \label{sec:app-robustness-Z}
To gain further insights into the mechanisms that define parental support $Z$, we conduct another robustness check that relaxes the definition of superior parental support (described in Section 1.1.2). Instead of requiring that youth rate 3 to both questions for all parents and caregivers, our relaxed definition allows for at most one rating of 2 for one of the four possible parent/question combinations. This relaxation increases the size of the intervention ($Z=1$) group by 1297 youth but does not change the size of the non-intervention ($Z=0$) group. The new sample size is $N = 5807$. The results from the sensitivity analysis are shown in Table \ref{tbl:abcd-robustness-z}.

\begin{table}[ht] 
 \centering 
 \begin{tabular}{lcccc} \toprule 
  \textbf{Parameter} & \textbf{Estimate (SD)} & \textbf{95\% CI} & $\Lambda^{\ast}_{0.05}$ & $\Lambda^{\ast}$ \\ \midrule 
 Observed Disparity & 0.234 (0.017) & [0.201, 0.268] & -- & --\\ 
 Disparity Reduction & 0.021 (0.017) & [-0.01, 0.05] & 1.00 & 1.05\\ 
 Residual Disparity & 0.213 (0.024) & [0.169, 0.260] & 1.54 & 1.69\\ \bottomrule 
 \end{tabular} 

 \caption{Results from robustness check using a more relaxed definition of superior parental support. Observed disparity, disparity reduction, residual disparity, and critical sensitivity parameters and 95\% confidence intervals are presented for the ABCD Study. $\Lambda^{\ast}_{0.05}$ corresponds to the critical parameter where the \textit{confidence interval} crosses 0, and $\Lambda^{\ast}$ corresponds to the critical parameter where the \textit{point estimate} crosses 0. There are no critical sensitivity parameters for the observed disparity since it is not a causal estimand.} 
 \label{tbl:abcd-robustness-z} 
\end{table} 

Using the relaxed definition of parental support shows similar patterns as before, except the disparity reduction is no longer statistically significant at level $\alpha = 0.05$. %, confirming our previous assertion that the result is sensitive to unmeasured confounding. 
This finding is expected: relaxing the parental support criteria lowers the effective ``dose" of parental support received by those in the treated group, and as such we expect a smaller intervention effect on average than in our original analysis.  % resulting in less disparity reduction. Conversely, defining parental support to include the highest-rated parents is likely to result in better mental health and less suicidality for sexual minority youth, hence the larger disparity reduction. 
While this relaxed definition broadens the inclusion criteria, it underlines our original finding that parental support explains at most a minor part of the disparity in suicidal ideation for sexual minority youth.

This small and insignificant disparity reduction creates an ideal setting to perform the three-sided test, as discussed in Section \ref{sec:equiv-test}. Based on the calibration plot, there would need to exist an unobserved confounder that is over five times as prognostic as family conflict (the strongest confounder) in order to mask a 100\% reduction in disparity, a scenario that is difficult to argue. The argument in the previous paragraph about the criteria for superior parental support appears to be a stronger argument for the presented results, indicating that further intervention screening is needed to identify strong targets that can reduce suicidality in sexual minority youth.  

Figures \ref{fig:reduction-abcd-robust-z} and \ref{fig:residual-abcd-robust-z} provide the calibration plots for disparity reduction and residual disparity, respectively. Note that there is no gray region for the disparity reduction plot since the result was not statistically significant.

\begin{figure}[ht]
    \centering    
    \includegraphics[width=0.75\linewidth]{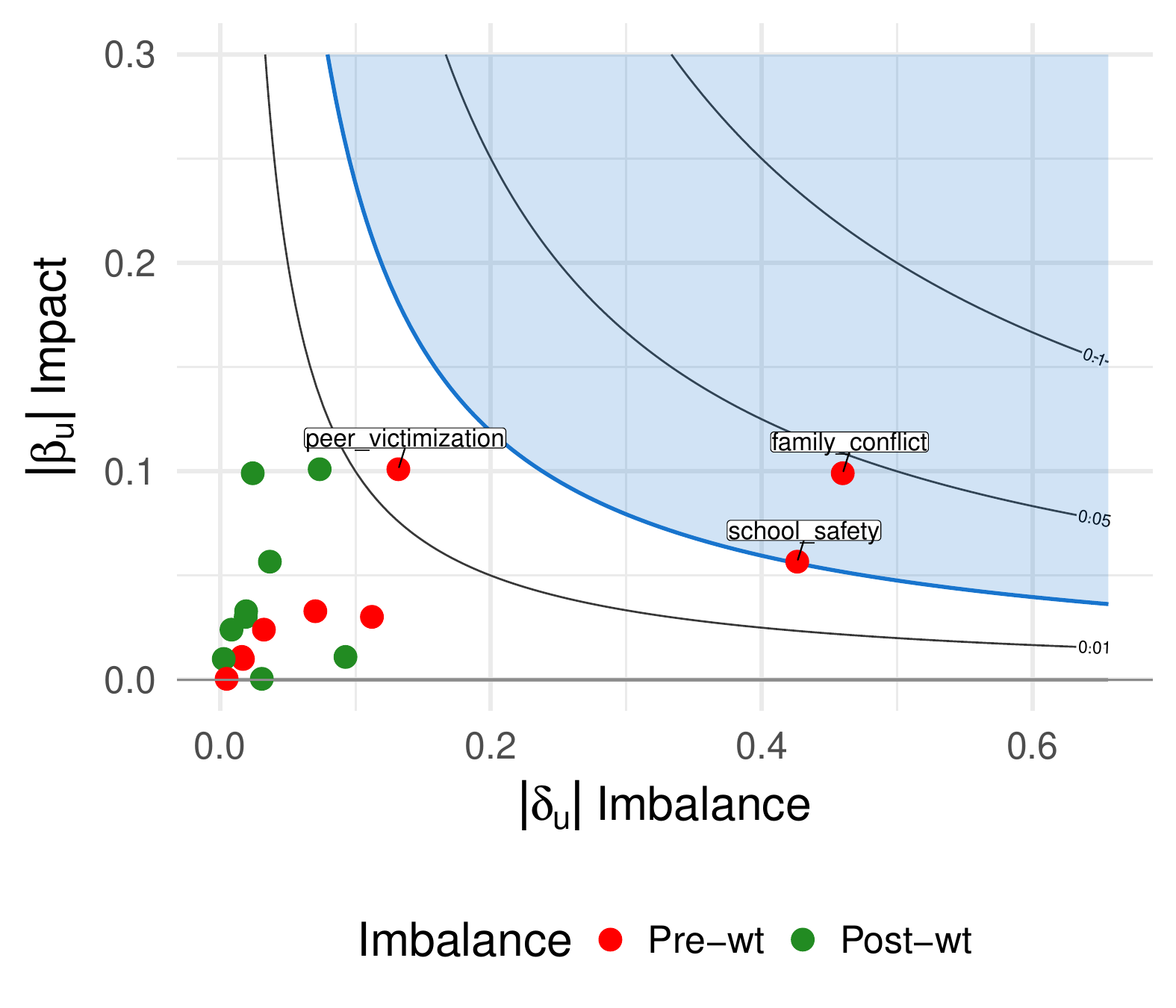}
    \caption{Bias contour plot of {\textbf{disparity reduction}} after relaxing definition of parental support. Note there is no gray area denoting the region where the confidence interval crosses 0 because the disparity reduction is already statistically insignificant. Please refer to Figures \ref{fig:reduction-abcd-contour} and \ref{fig:residual-abcd-contour} for plot details.}
    \label{fig:reduction-abcd-robust-z} 
\end{figure}
\begin{figure}[ht]
    \centering    
    \includegraphics[width=0.75\linewidth]{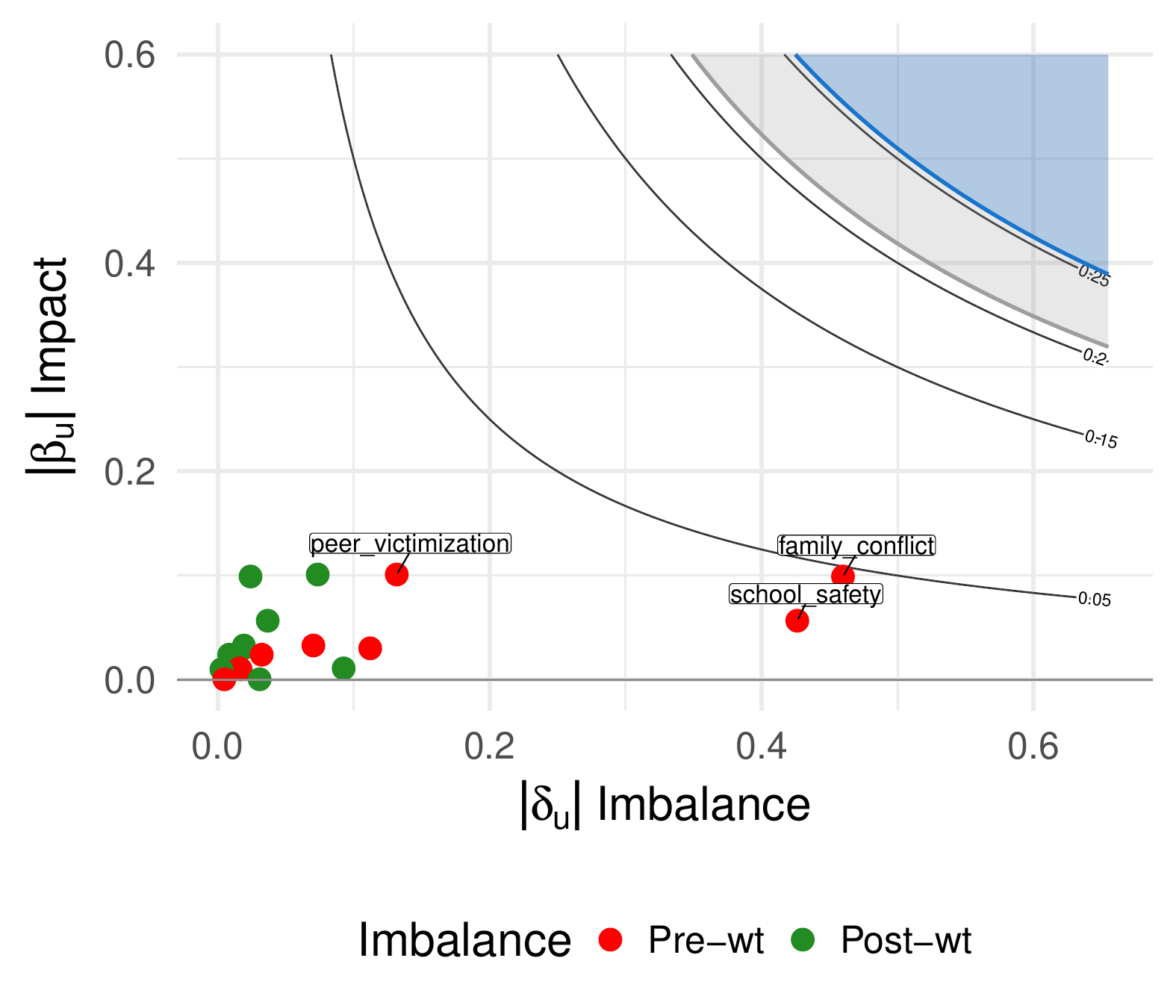}
    \caption{Bias contour plot of {\textbf{disparity reduction}} after relaxing definition of parental support $Z$. Please refer to Figures \ref{fig:reduction-abcd-contour} and \ref{fig:residual-abcd-contour} for plot details.}
    \label{fig:residual-abcd-robust-z} 
\end{figure}

\end{revisiontwo}
\end{document}